\documentclass{aa} 

\DeclareRobustCommand{\rchi}{{\mathpalette\irchi\relax}}
\newcommand{\irchi}[2]{\raisebox{\depth}{$#1\chi$}} 

\newcommand{\er}{\mathbf{e}_{\rm r}}
\newcommand{\ex}{\mathbf{e}_{\rm x}}
\newcommand{\ey}{\mathbf{e}_{\rm y}}
\newcommand{\ez}{\mathbf{e}_{\rm z}}

\newcommand{\rlight}{r_{\rm L}}

\newcommand{\etheta}{\mathbf{e}_\vartheta}

\usepackage{graphicx}
\usepackage{txfonts}

\begin{document} 

\title{Multi-wavelength pulse profiles from the force-free neutron star magnetosphere}

\author{
J. P\'etri\inst{1}
}

\institute{
Universit\'e de Strasbourg, CNRS, Observatoire astronomique de Strasbourg, UMR 7550, F-67000 Strasbourg, France.\\
\email{jerome.petri@astro.unistra.fr}         
}

\date{Received ; accepted }

 
  \abstract
   {The two last decades have witnessed dramatic progresses in our understanding of neutron star magnetospheres thanks to force-free and particle in cell simulations. However, the associated particle dynamics and its emission mechanisms and locations are not fully constrained notably in X-rays.}
   {In this paper, we compute a full atlas of radio, X-ray and $\gamma$-ray pulse profiles relying on the force-free magnetosphere model. Our goal is to use such data bank of multi-wavelength profiles to fit a substantial number of radio-loud $\gamma$-ray pulsars also detected in non-thermal X-rays to decipher the X-ray radiation mechanism and sites. Using results from the third $\gamma$-ray pulsar catalogue (3PC), we investigate the statistical properties of this population.}
   {We assume that radio emission emanates from field lines rooted to the polar caps, at varying height above the surface, close to the surface, at an altitude about 5-10\% of the light-cylinder radius~$\rlight$. The X-ray photons are produced in the separatrix region within the magnetosphere, i.e. the current sheet formed by the jump from closed to open magnetic field lines. We allow for substantial variations in emission height. The $\gamma$-ray are produced within the current sheet of the striped wind, outside the light-cylinder. }
   {A comprehensive set of light-curves in radio, X-ray and $\gamma$-ray has been computed. Based on only geometric considerations about magnetic obliquity, line of sight inclination and radio beam cone opening angle, pulsars can be classified as radio-loud or quiet and $\gamma$-ray loud or quiet. We found that the 3PC sample is compatible with an isotropic distribution of obliquity and line of sight.}
   {The atlases constructed in this work are the fundamental tools to explore individual pulsars and fit their multi-wavelength pulse profiles in order to constrain their magnetic topology, the emission sites as well as the observer line of sight.}


\maketitle

%

\section{Introduction}

Nowadays, almost 300~$\gamma$-ray pulsars are reported in the Fermi pulsar catalogue\footnote{297 are announced on \url{https://confluence.slac.stanford.edu/display/GLAMCOG/Public+List+of+LAT-Detected+Gamma-Ray+Pulsars}.} but see also the third pulsar catalogue detailed in \cite{smith_third_2023}. Among them, more than 20 were detected pulsating simultaneously in radio and X-ray. This population of pulsars is a perfect target to study the multi-wavelength emission properties of neutrons stars. The complete list of radio pulsars has been compiled in the ATNF Pulsar Catalogue of \cite{manchester_australia_2005} and is kept up to date online. Unfortunately, no exhaustive non-thermal X-ray pulsar catalogue exist to date. However some recent papers like \cite{cotizelati_spectral_2020} acknowledge the simultaneous detection of non-thermal X-rays and $\gamma$-rays of a significant sample of 40 pulsars. \cite{chang_observational_2023} found 32 pulsars described by a power-law (PL) + black body (BB) model. A single PL model fits the other 36 pulsars well (thus 68 in total). These numbers are however only lower limits because very recently \cite{mayer_searching_2024} identified several dozen possible X-ray pulsars of the unassociated Fermi-LAT sources with SRG/eROSITA (Spektrum-Roentgen-Gamma), about 30-40 candidates distributed equally in young and recycled pulsars. In the higher energy band of soft $\gamma$-rays, another useful list of pulsars is given by \cite{kuiper_soft_2015}. Phase-aligned multi-wavelength pulsar light curves are the milestones to constrain the location of radio, X-ray and $\gamma$-ray emission within the magnetosphere and near wind as attempted since several decades.

One of the most promising joint radio and $\gamma$-ray model relies on the pulsar striped wind geometry. This idea has been investigated in depth the last two decades starting from \cite{kirk_pulsed_2002}. It was then applied to the Fermi $\gamma$-ray pulsars by \cite{petri_unified_2011} in the split monopole picture, or more recently by \cite{benli_constraining_2021} and \cite{petri_young_2021} in the force-free dipole magnetosphere. These works showed that a joint radio+$\gamma$-ray pulse profile fitting can drastically constrain the magnetic dipole moment inclination angle with respect to the rotation axis as well as the line of sight inclination angle. These fitting were based on numerical quantitative simulations of the force-free magnetosphere performed by \cite{petri_pulsar_2012}.
These kind of numerical simulations were pioneered by \cite{contopoulos_axisymmetric_1999} for the axisymmetric magnetosphere and extended by \cite{spitkovsky_time-dependent_2006} to an oblique rotator. However strictly speaking, the force-free model does not radiate because it is dissipationless as no particle acceleration occurs. Therefore the paradigm shifted to a kinetic description of these magnetospheres including single particle acceleration in 3D \citep{philippov_ab_2015} and radiation feedback self-consistently to predict observational signatures \citep{cerutti_modelling_2016} that even predict the polarization properties \citep{cerutti_polarized_2016}. The striped wind and its dissipation was also investigated by PIC simulations in \cite{cerutti_dissipation_2017, cerutti_dissipation_2020}. Sticking closer to $\gamma$-ray observations, \cite{brambilla_electronpositron_2018}, \cite{kalapotharakos_fermi_2017} and \cite{kalapotharakos_three-dimensional_2018} performed similar simulations.

Several other emission regions were proposed in the past. The outer gaps, the slot gaps and the polar caps are the favourite places for pulsed emission. They predict distinct pulse profiles helping to discriminate between competing models \citep{dyks_two-pole_2003, dyks_relativistic_2004}. Producing exhaustive atlases of light curves in all relevant wavelength is the starting point for any broadband fitting task. For instance \cite{watters_atlas_2009} focused on young $\gamma$-ray pulsar profiles and \cite{pierbattista_light-curve_2015, pierbattista_young_2016} on different emission mechanisms for a significant sample of pulsars whereas \citet{johnson_constraints_2014} focused on millisecond pulsars. Two-pole caustics and outer gap atlases were presented in \citet{harding_gamma-ray_2011}.

The third Fermi $\gamma$-ray pulsar catalogue \citep{smith_third_2023} summarizes all our knowledge about $\gamma$-ray pulsars spectra and light-curves. Because force-free computations do not catch the acceleration and radiation mechanism within the magnetosphere and wind, resistive or dissipative magnetospheres where built, like force-free inside dissipative outside (FIDO) as introduced by \cite{kalapotharakos_toward_2012}, and used by \citet{brambilla_testing_2015} for detailed pulse $\gamma$-ray investigations. For some reviews on the interplay between magnetospheric modelling and $\gamma$-ray observations see \cite{harding_gamma-ray_2016} and \cite{venter_high-energy_2018}.

Whereas more than 3000~pulsars are known to be radio emitters, only 10\% of them are seen shining in $\gamma$-rays. The release of the third pulsar catalogue by \cite{smith_third_2023} furnishes a wealth of new and accurate data helping in our task to decipher the high-energy radiation mechanism and location. Although many studies focused on the radio and $\gamma$-ray emission, using fluid or particle simulations \citep{cerutti_modelling_2016, kalapotharakos_fermi_2017, kalapotharakos_gamma-ray_2023, watters_atlas_2009, pierbattista_young_2016}, none really addressed the pulsed X-ray emission properties. This article fills the gap by exploring quantitatively the X-ray light-curves morphology depending on the pulsar geometry.

Sec.~\ref{sec:site_emission} reminds the emission location of the different energy bands: radio, X-ray and $\gamma$-rays. Next Sec.~\ref{sec:multi_lambda} summarizes the light-curves in these respective wavebands. Sec.~\ref{sec:Energetics} discusses the energetics of the radiating particles by computing the curvature of field lines and the associated photon energy and particle dynamics. Some statistical expectations are compared to the current observational status in Sec.~\ref{sec:Statistics}. Conclusions and future perspectives are presented in Sec.~\ref{sec:Conclusions}.

\section{Emission sites}
\label{sec:site_emission}

In the last years, three standard emission sites emerged from different studies of the pulsar magnetosphere. The first one is the polar cap region, close to the stellar surface and responsible for the coherent radio emission. The second one follows the region along the separatrix, i.e. the transition surface between close and open magnetic field lines within the light-cylinder. The third one radiates in the current sheet of the pulsar striped wind. We briefly remind the main features of such emission regions that have been introduced in a similar way by \cite{petri_general-relativistic_2018}, including general-relativistic effects.

The setup assumes that the magnetic dipole moment $\vec{\mu}$ is directed along a unit vector~$\vec{m}$ such that $\vec{\mu} = \mu \, \vec{m}$. This magnetic moment rotates at an angular speed~$\Omega$ around the rotation axis directed along the vector~$\ez$ of the Cartesian basis vectors $(\ex,\ey,\ez)$. The observer line of sight unit vector points towards the direction~$\vec{n}$. The emitting particle density number $n$ is assumed to be spherically symmetric and to decrease like a power law in spherical radius~$r$ with an exponent $q$ such that
\begin{equation}\label{eq:densite}
	n(r) \propto r^{-q} .
\end{equation}
Particle density number is therefore prescribed, we set $q=2$ in all wavelengths. The Lorentz factor of the moving particles is set to $\Gamma=10$.

\subsection{Magnetic field geometry}

\begin{figure}
	\centering
	\includegraphics[width=0.9\linewidth]{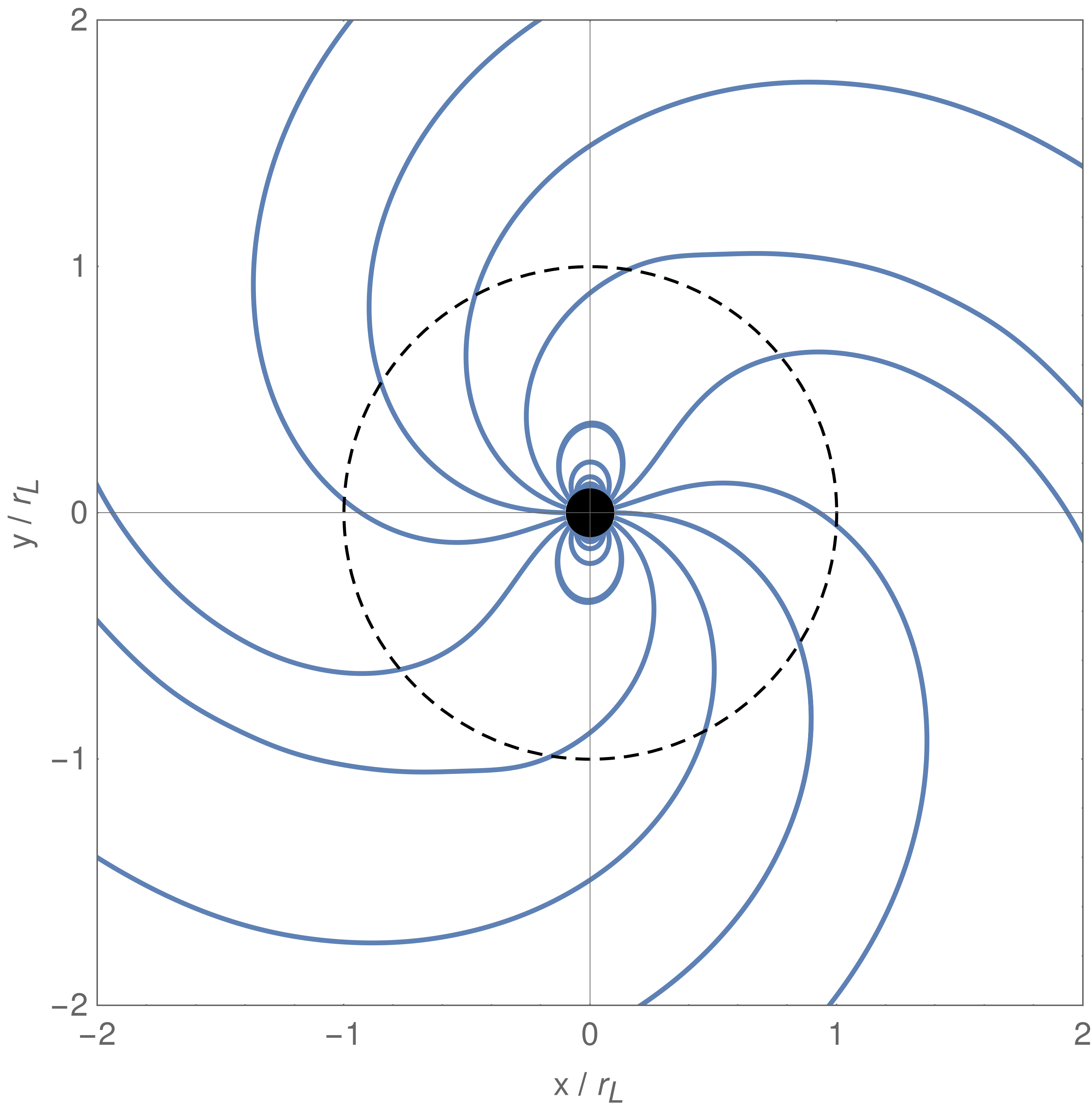}
	\caption{Magnetic field lines in the equatorial plane for an orthogonal rotator with $R/\rlight=0.1$. The neutron star is shown as a black disk and the light cylinder as a dashed black circle.\label{fig:lignes_equatoriales}}
\end{figure}

The structure of the magnetic field lines fully determines the size and location of the polar cap, the geometry of the separatrix surface and the location of the current sheet within the striped wind. These are the three sites producing respectively radio photons, X-rays and $\gamma$-rays. Our model is based on numerical simulations of the neutron star magnetosphere using spectral methods in the force-free approximation, updating previous computations performed by \cite{petri_pulsar_2012} and \cite{petri_young_2021}. The ratio between the neutron star radius~$R$ and the light-cylinder radius~$\rlight$ is set to $R/\rlight=0.1$ corresponding to the altitude where radio emission is expected to be produced. 
Close to the surface, the magnetic field is reminiscent to a static dipole corotating with the star whereas outside the light-cylinder magnetic field lines open up and form a current sheet wobbling around the equatorial plane. An example of field lines in the equatorial plane for an $\rchi=90\degr$ obliquity is shown in figure~\ref{fig:lignes_equatoriales}.

\subsection{Polar cap and radio}

A careful analysis of emission height in normal pulsars based on the rotating vector model and using a large sample of radio pulsars \citep{weltevrede_profile_2008, mitra_nature_2017, johnston_period-width_2019, johnston_thousand-pulsar-array_2023} showed that radio emission is emanating from regions well above the polar caps, around 5\% to 10\% of the light cylinder radius. For pulsars showing interpulses, the geometry is even better constrained because they are close to orthogonal rotators \citep{johnston_beam_2019}. We use this proxy to implement a simple radio pulse profile emissivity that drops as a Gaussian when moving away from the magnetic axis. The width of the pulse profile is defined by the typical half-opening angle of the emission cone at an altitude~$h_{\rm e}$ above the stellar surface 
\begin{equation}\label{eq:radio_cone_opening}
	\rho \approx \frac{3}{2} \, \theta_{\rm pc}
\end{equation}
with half-opening angle of the cone locating the root of the field lines as
\begin{equation}\label{eq:field_cone_opening}
	\sin \theta_{\rm pc} = \sqrt{\frac{h_{\rm e}}{\rlight}} .
\end{equation}
The emissivity within the radio emitting region and related to the Gaussian beam shape reads
\begin{equation}\label{eq:Emissivity_Radio}
 \epsilon_r \propto n(r) \, e^{-3\,(\theta/\theta_{\rm pc})^2} .
\end{equation}
Radio photons are assumed to be produced at a small emission height interval starting from the inner boundary of the simulation box $h_{\rm e}=R_1$. In the present simulations this is set to $R_1/\rlight=0.1$ thus 10\% of the light-cylinder radius as found in the above-mentioned literature. The extension in radius is set to 1\% of $\rlight$ amounting to a height range $[0.1,0.11]~\rlight$. As the magnetic moment $\vec{m}$ rotates, the dot product $\vec{m} \cdot \vec{n}$ evolves with time according to $\Omega\, t $ and reaches its maximal value whenever $\ez$, $\vec{m}$ and $\vec{n}$ are coplanar. This occurs whenever the north or south pole points towards the observer. Photons are emitted along the local particle velocity vector $\vec{v}$ which is a combination of corotation velocity and velocity along magnetic field lines such that
\begin{equation}\label{eq:vitesse}
 \vec{v} = \vec{\Omega} \wedge \vec{r} + f \, \vec{B} .
\end{equation}
The unknown parameter~$f$ is found by imposing the Lorentz factor~$\Gamma$ of the particle all other quantities being known.
Expression~\eqref{eq:vitesse} takes into account the aberration effect through the corotating velocity field in the first term of the right hand side. Retardation are including through the time of flight delay of photons. These aberration/retardation effects are discussed in \cite{blaskiewicz_relativistic_1991}.

\subsection{Separatrix and X-rays}

Farther away from the radio emitting region, in the separatrix zone where the transition between close and open magnetic field lines occurs, we expect X-rays photons to be produced from the synchrotron and/or curvature radiation mechanism. 
This X-ray component is produced by the secondary electron-positron pairs radiating synchrotron photons. For instance \cite{kisaka_synchrotron_2014} computed the synchrotron emission for old pulsars to explain the non-thermal X-ray emission whereas \cite{kisaka_efficiency_2017} investigated a similar mechanism to estimate the efficiency of this emission in rotationally powered pulsars. \cite{takata_x-ray/gev_2017} discussed the connection between X-ray and GeV emission in this framework. \cite{torres_synchrocurvature_2019} showed that the synchro-curvature radiation is able to explain the non-thermal X-ray spectra of several pulsars but they did not compute pulse profiles, see also some refinement by \cite{iniguez-pascual_synchro-curvature_2022}. For polar cap and outer gap scenarios of non-thermal X-ray pulsed emission see also \cite{peng_non-thermal_2008} and \cite{zhang_non-thermal_2006}. In a narrow layer of small thickness parametrized by $\sigma_{\rm sep}$ (much less than the light-cylinder radius) and delimited by magnetic flux tubes, with a constant emissivity along field line but decaying like a Gaussian for field lines moving away from the separatrix surface, X-ray photons start propagating tangentially to the local direction of the velocity vector eq.~\eqref{eq:vitesse} where they were generated. More precisely the emissivity of a given point attached to a field line  reads
\begin{equation}\label{eq:Emissivity_X}
	\epsilon_x \propto n(r) \, e^{-\ell^2_{\rm min}/\sigma_{\rm sep}^2} 
\end{equation}
where $\ell_{\rm min}$ is the minimal distance of this field from the light-cylinder. By construction any point on this field line will have the same emissivity as another point on the same field line except for a modulation by the particle density number~$n(r)$.
The location of the X-ray emission sites are largely unconstrained so far. In order to get an idea of the emission height and extension, we divided the separatrix from the inner boundary simulation box to the light-cylinder into a set of uniform intervals of radial length $0.1\rlight$ from $r=R_1$ to $r=\rlight$. As the radiation is additive, any realistic light-curve is given by summing several adjacent regions to construct the total light-curve from the individual building block light-curves. In this manner, we deduce the lower and the upper emission heights along the separatrix surface.

\subsection{Striped wind and $\gamma$-rays}

The most energetic photons are produced in the current sheet of the striped wind, outside the light-cylinder. Emission starts straight from the light-cylinder up to several $\rlight$ with an emissivity decaying with distance due to the decreasing particle density number~$n(r)$.
The emitting current sheet is supposed to be infinitely thin and located in space where the radial component of the magnetic field changes sign between the upper part $B_r^+$ (connected to the north pole) and the lower part $B_r^-$ (connected to the south pole) such that $B_r^+ \, B_r^-<0$. The locus thus formed corresponds to a surface like a ballerina skirt similar to the solar wind structure. The emissivity therefore vanishes out of the surface defined by the current sheet and equals
\begin{equation}\label{eq:Emissivity_Gamma}
	\epsilon_\gamma \propto n(r)
\end{equation}
on this surface. The emission starts outside the light-cylinder at cylindrical distances $r\sin\theta\geq \rlight$ and the wind propagates radially at a Lorentz factor~$\Gamma$ fixed to $\Gamma=10$. No other free parameter is required in this minimal model.
The base of the radiating wind $r_{\rm in}$ is located outside the light-cylinder and extends to a spherical radius $r_{\rm out} > r_{\rm in}$. This outer boundary is not constrained but the wind emissivity decreases with the distance~$r$ and the current sheet geometry slightly changes, based on the magnetic topology. At large distances, its luminosity becomes negligible so we arbitrarily cut the wind at a variable radius~$r_{\rm out}$ in order to quantify the impact of the emission height in $\gamma$-rays onto the light curves. Therefore we defined two intervals, a first in $[1,2]\rlight$ and a second in $[2,3]\rlight$, comparing the evolution of the light-curves with distances.

\section{Multi-wavelength radiation patterns}
\label{sec:multi_lambda}

In this section we discuss in depth the multi-wavelength pulse profile characteristics and evolution with the geometry controlled by the obliquity $\rchi$ and the line of sight inclination angle $\zeta$. First we summarize all the results by plotting intensity maps in radio, X-ray and $\gamma$-ray. Next we show typical atlases of light-curves by varying $\rchi$ and $\zeta$. Finally, for a given geometry with fixed values of $\rchi$ and $\zeta$, we investigate the pulse shape dependence on the emission height, especially in X-rays where these variations with altitude are the most prominent.

\subsection{Geometric constraints}

Due to the particular configuration of the polar cap and the striped wind emission, it is possible to derive simple relations between the obliquity and the line of sight inclination to decide if radio and/or $\gamma$-ray photons are detected or not. For X-ray detection, the criteria is more difficult to derive.

The current sheet of the striped wind crosses the observer line of sight whenever the latter lies close to the equatorial plane, that is
\begin{equation}\label{eq:zeta_gamma_visible}
\left| \zeta - \frac{\pi}{2} \right| \leq \rchi .
\end{equation}
This condition is derived in \cite{petri_unified_2011} for the split monopole solution that does not significantly differ from the dipole force-free model obtained by numerical solution. This condition is very general as it does not depend on the exact nature of the field line geometry inside the light-cylinder as long as the dominant multipole becomes the dipole at the light-cylinder.
For a radio emission cone of half opening angle $\rho$ the visibility condition reads for one pole
\begin{equation}\label{eq:zeta_radio_visible_nord}
\left| \zeta - \rchi \right| \leq \rho
\end{equation}
whereas for the other pole it becomes
\begin{equation}\label{eq:zeta_radio_visible_sud}
\left| \zeta + \rchi - \pi \right| \leq \rho.
\end{equation}
This is due to the symmetry between the angle $\rchi$ and $\pi-\rchi$ when considering the emission patterns of the magnetosphere. To observe two radio pulses, both conditions must be satisfied simultaneously leading to the constrain
\begin{equation}\label{eq:zeta_radio_visible}
	\left| \zeta - \frac{\pi}{2} \right| \leq \rho .
\end{equation}
In other words, the line of sight must not deviate from the equatorial plane more than the radio emission cone half opening angle. A second conditions on the obliquity must be met 
\begin{equation}\label{eq:chi_radio_visible}
\left| \rchi - \frac{\pi}{2} \right| \leq 2 \, \rho
\end{equation}
meaning that it must be almost an orthogonal rotator, within an interval equal to twice the radio cone half opening angle.
The value of $\rho$ is directly related to the emission height $h_{\rm e}$ as given by eq.\eqref{eq:radio_cone_opening}. Radio observations suggest that $(h_{\rm e})/\rlight\lesssim0.1$ thus $\rho \lesssim 27.6\degr$.

\begin{figure}[h]
	\centering
\includegraphics[width=0.7\columnwidth]{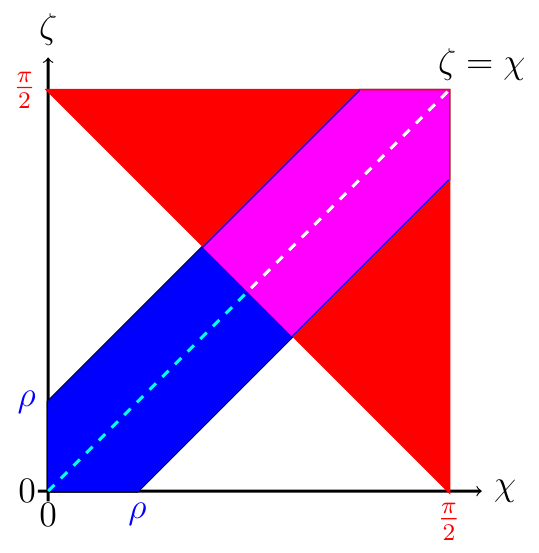}
\caption{Geometry showing the region of radio and $\gamma$-ray visibility depending on the magnetic obliquity~$\rchi$ and line of sight~$\zeta$. The red area corresponds to $\gamma$-ray only, the blue area to radio-only with $\rho=20\degr$ and the magenta area to radio-loud $\gamma$-ray visibility. Due to the symmetry of the plot, the three other quadrants with $\rchi\geq\pi/2$ or $\zeta\geq\pi/2$ are not shown.\label{fig:contraintes}}
\end{figure}

Figure~\ref{fig:contraintes} summarizes the constraints about radio and $\gamma$-ray visibility depending on~$\rchi$ and $\zeta$. The red area corresponds to $\gamma$-ray-only, the blue area to radio-only and the magenta area to radio-loud $\gamma$-ray visibility. The observer looks exactly onto the magnetic axis when the geometry falls onto one of the two dashed diagonals $\zeta=\rchi$ or $\zeta=\pi-\rchi$. The pattern is highly symmetric due to the north-south symmetry of the magnetic field and due to the stellar rotation. For instance $ \rchi=\pi/2$ and $\zeta=\pi/2$ are symmetry axes.

For the X-ray constraints, finding a simple analytical condition for visibility is impossible because the exact geometry of the separatrix is not known in the force-free model. However to make progress in this direction and derive an approximate expression for detecting pulsed X-ray, we assume that the photons emanate from well within the light-cylinder at distances $r\lesssim \rlight$. In this region, the magnetic field remains dipolar to an acceptable precision and the polar cap size and shape still resemble the vacuum case although deviations are observed as has been checked in several previous works. For recent results see for instance the figure~11 of \cite{petri_radiative_2022} where quantitative comparisons have been done. We therefore use the static dipolar separatrix structure as a proxy to predict the pulsed X-ray detectability. To study the impact of the dipolar geometry, we use dipolar coordinates as detailed in the appendix~\ref{app:A} and used by \cite{swisdak_notes_2006,orens_vector_1979}. In this dipole approximation, the condition for pulsed X-ray detection mimics the radio visibility condition, the only difference being the size of the emitting cone because of the varying emission altitude postulated in X-rays and the fact that the cone is hollow. However, because of the stellar rotation even a hollow cone will cross the line of sight when the condition
\begin{equation}
	\left| \zeta - \rchi \right| \leq \rho_{\rm x}
\end{equation}
is fulfilled, $\rho_{\rm x}$ being the maximum X-ray emission cone half-opening angle. It depends on the boundaries of the emitting region located between the radius $h_1$ and $h_2$. Actually, because $\theta_{\rm x}$ is a monotonic function of $h_{\rm e}$ it increases with $h_{\rm e}$ and the maximum is given by
\begin{equation}
	\cos \rho_{\rm x} = \sqrt{1-h_2/\rlight}
\end{equation}
the cone opening angle at lower altitude being always smaller.

\subsection{Maps}

\begin{figure*}[h]
	\centering
	\includegraphics[width=\linewidth]{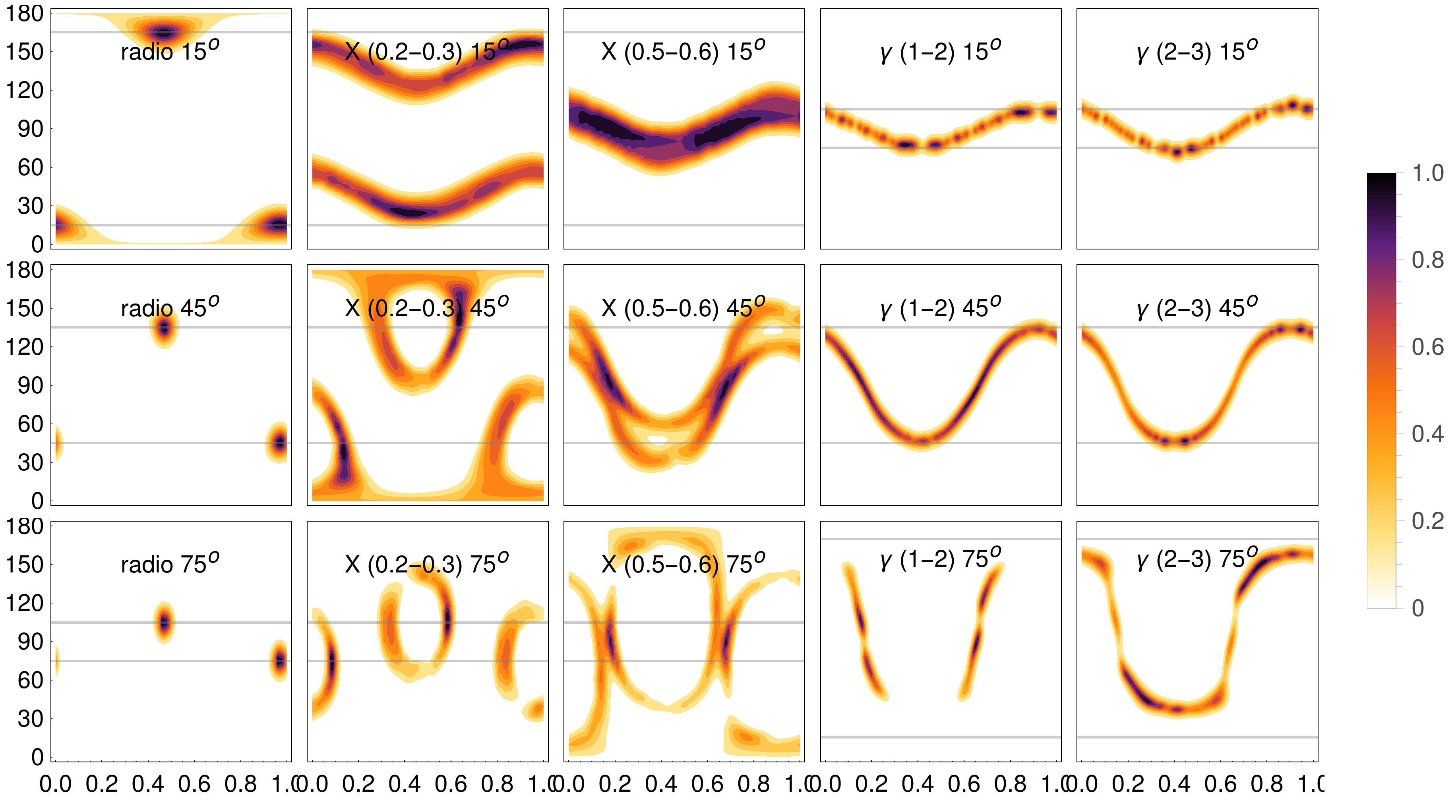}
	\caption{Map of radio, X-ray and $\gamma$-ray radiation for different emission altitudes and for $\rchi=\{15\degr,45\degr,75\degr\}$. The legend shows the wavelength, the emission interval and the obliquity.
	\label{fig:carte_complete_gamma_r0.2}}
\end{figure*}

Figure~\ref{fig:carte_complete_gamma_r0.2} shows the multi-wavelength maps of radio, X-ray and $\gamma$-ray emission. On the left column, the radio emission is produced in the range $r/\rlight \in[0.1,0.11]$, on the second column, the non-thermal X-ray from an emission height in $r/\rlight\in[0.2,0.3]$ whereas the third columns assumes the same emission in $r/\rlight\in[0.5,0.6]$, the last two columns the $\gamma$-ray emission emanating from the interval $r/\rlight\in[1,2]$ and from the interval $r/\rlight\in[2,3]$. We observe the archetypal S-shape pattern produced by the striped wind structure. The high-energy emission pattern is not significantly sensitive to the emission height, as long as the radiation sites remain close to the light-cylinder. The contribution for large distances are smeared out and decay quickly with radius due to the density dependence $n(r)$. For $r\gtrsim3-4\rlight$ the emission produced becomes insignificant without any impact on the light-curves thus it was discarded.

At low altitudes $r/\rlight\in[0.1,0.2]$, the X-ray emission pattern looks like an oval encircling the radio emission pattern. As the height increases, this pattern approaches the S-shape pattern of the striped wind. The dependence on altitude is important for X-ray because it follows the drastically changing magnetic field topology within the light-cylinder, from an almost static dipole configuration close to the surface to a retarded dipole close to the light-cylinder and transitioning to an electromagnetic wave outside it.

The above maps give a full representation of the whole set of light-curves being produced in the different emission sites. In the next paragraph we show atlases of radio, X-ray and $\gamma$-ray light-curves to be used in fitting multi-wavelength pulse profile of radio-loud $\gamma$-ray pulsars also detected in non-thermal X-rays. A detailed example will be discussed in another paper.

\subsection{Light curves}

All possible radio pulse profiles from a centred dipole are shown in the appendix, figure~\ref{fig:atlas_radio_r0.2}. For an almost aligned rotator, the pulsed fraction is weak at all line of sight inclination angles~$\zeta$ and the pulsation could be hardly detectable depending on the noise level. When far from an orthogonal rotator, only one pulse is visible from one magnetic pole. Only for $\zeta\approx90\degr$ are both pulses visible and separated by half a period, due to the symmetry of the dipole field. The detection of a double pulse structure depends on the angles $(\rchi, \zeta)$ but also on the beam width. In the special case of $\zeta=90\degr$, both pulses are symmetrical and always detected whatever $\zeta$. For lower obliquities $\zeta \lesssim 90\degr$, one pulse dominates, see for instance the case $\{\rchi,\zeta\} = \{75\degr,80\degr\}$.

The $\gamma$-ray light-curves are shown in figure~\ref{fig:atlas_gamma_r1}. Inspecting this figure with $r/\rlight\in[1,2]$, the pattern shows one peak or two peaks depending on the peculiar combination $\{\rchi,\zeta\}$. It behaves similarly to the radio pulse profile. However, in the $\gamma$-ray band, there is much more chance to produce a double-peaked light-curve because this configuration is much more favourable geometrically. A double peak is observed whenever $|\zeta-\pi/2|\leq\rchi$. Moving to higher altitudes in the range $r/\rlight\in[2,3]$ does not alter these findings, figure~\ref{fig:atlas_gamma_r1}.

The new part of this work concerns the X-ray radiation, supposed to be produced along the separatrix, on an unspecified altitude that is left as a free parameter. At the lowest altitude, in the range $r/\rlight\in[0.1,0.2]$, the pulse profiles look like in figure~\ref{fig:atlas_X_r0.2}. We observe between one and four pulses depending on the geometry. A single pulse if observed for low obliquity and low line of sight inclination, upper left corner of the plot, whereas four pulses are observed at high obliquity (close to an orthogonal rotator) and high line of sight inclination angles. At higher altitudes, for instance in the range $r/\rlight\in[0.4,0.5]$, the four peaks transform into a double peak profile, figure~\ref{fig:atlas_X_r0.5}.

We plan to use these X-ray atlases to constrain the emission location of the X-ray photons. To this end, we assume that the pulsar obliquity is well determined by the joint radio and $\gamma$-ray pulse profile fitting, extracting accurate values for both angles $\rchi$ and $\zeta$. Then, to estimate the X-ray emission height, it is sufficient to explore a small subset of the entire X-ray atlas. As an example we show the evolution of the pulse profile with minimum and maximum altitudes $h_1$ and $h_2$ in figure~\ref{fig:courbelumierexextension} for a pulsar with obliquity $\rchi=45\degr$.

\section{Radiation energetics}
\label{sec:Energetics}

The geometric approach of fitting light-curves is not sufficient to to grasp all the dynamics of particle acceleration and radiation within the magnetosphere and wind. The energetic considerations of the whole electrodynamic processes is another important aspect of the problem that we discuss in this section.

\subsection{Curvature of field lines}

Well inside the light-cylinder, where the radio emission is produced, to a good accuracy, the magnetic field follows the static dipole geometry. To estimate quickly the curvature of field lines near the surface, along the separatrix and inside the radio emission beam, we start by computing the static dipole curvature. For an aligned dipole, each field line is labelled by a parameter $\lambda$ such that 
\begin{equation}\label{eq:field_lines}
 r(\lambda, \theta) = \lambda \, \sin^2 \theta
\end{equation}
$r$ being the radius and $\theta$ the azimuth in a spherical coordinate system. Along this particular field line symbolised by $\lambda$, the curvature radius $\rho_c$ depends on $\theta$ as
\begin{equation}\label{eq:rayon_courbure}
 \rho_c(\lambda, \theta) = \frac{\lambda}{3} \, \sin \theta \, \frac{(1 + 3 \cos^2 \theta)^{3/2}}{1 + \cos^2 \theta} .
\end{equation}
Along the magnetic axis, the curvature vanishes but as we will show, for the rotating dipole, the curvature does not vanish due to magnetic sweep back. By construction the separatrix is defined by field lines satisfying $\lambda=\rlight$. Close to the polar caps, the angle $\theta$ is small and the curvature tends to
\begin{equation}\label{eq:rayon_courbure_2}
\frac{\rho_c(\rlight, \theta)}{\rlight} \approx \frac{4}{3} \, \theta
\end{equation}
which at the stellar surface gives $\theta \approx \sqrt{R/\rlight}$ and therefore
\begin{equation}\label{eq:rayon_courbure_3}
\rho_c(\rlight, \theta) \approx \frac{4}{3} \, \sqrt{R\,\rlight} .
\end{equation}
The neutron star radius~$R$ being fixed to an almost constant value of $R\approx 12$~km, the curvature radius increases according to $\sqrt{\rlight} \propto \sqrt{P}$. The curvature radius increases with the pulsar period thus the frequency of curvature radiation diminishes right at the surface. In order to keep a typical radio photon range in the interval [10~MHz, 10~GHz], the emission altitude must increase too.

\begin{figure}[h]
	\centering
	\includegraphics[width=0.9\linewidth]{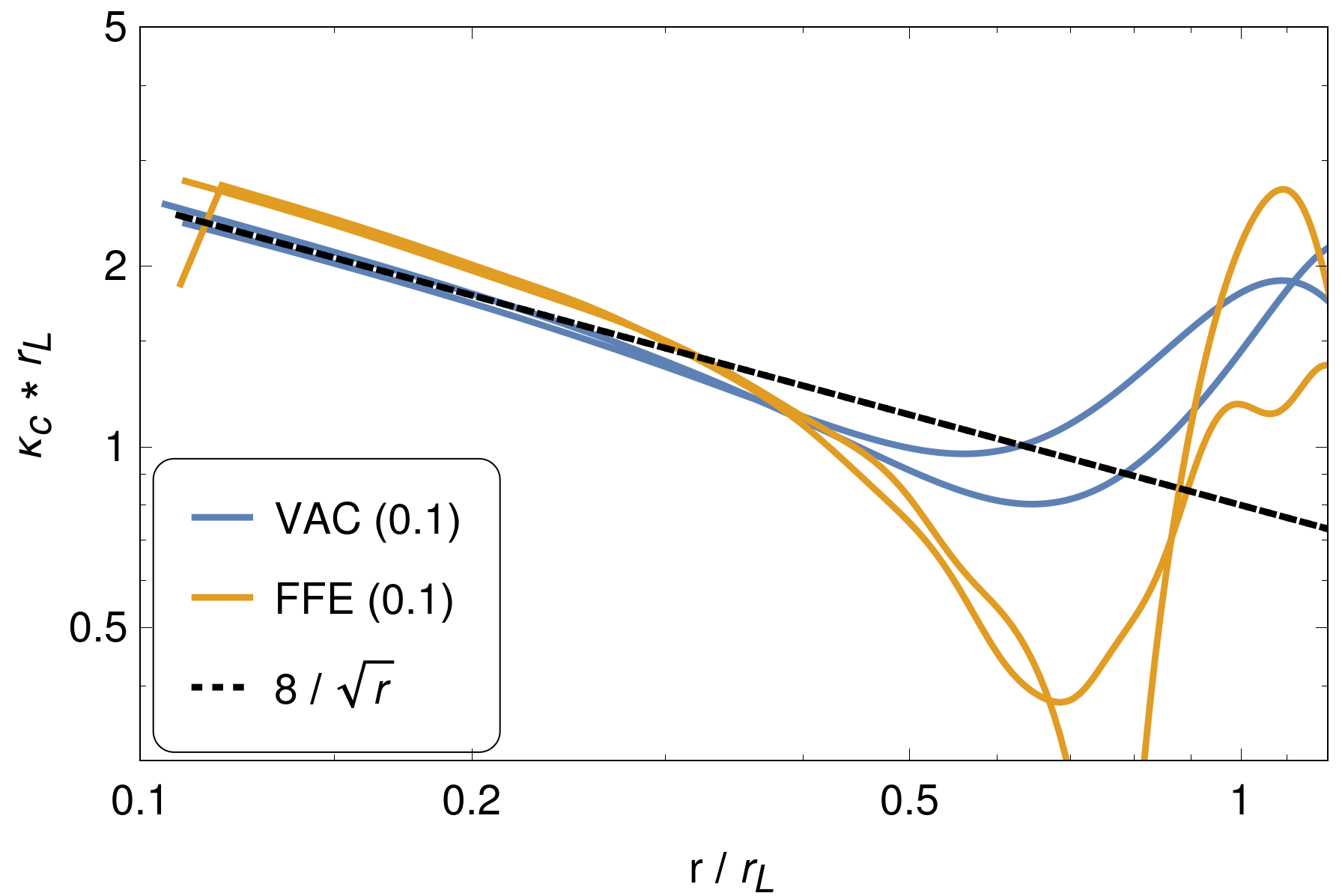}
	\caption{Curvature in units of $1/\rlight$ along the separatrix starting from the surface going to the light-cylinder and back to the surface for the vacuum field (VAC in orange) and the force-free solution (FFE in blue). Values are shown for an obliquity of $\rchi=75\degr$.}
	\label{fig:courbure_separatrice_vide_ffe}
\end{figure}

The curvature~$\kappa_c = 1 / \rho_{\rm c}$ along the separatrix is shown in Figure~\ref{fig:courbure_separatrice_vide_ffe}, in units of $1/\rlight$ for the vacuum Deutsch solution in blue (VAC) and for the force-free magnetosphere in orange (FFE). The inner boundary of the simulation box has been fixed to $R_1/\rlight=0.1$. The curvature decreases with distance to the star like $r^{-1/2}$ for the vacuum and approximately the same law $\sim r^{-1/2}$ for the FFE, as long as $r\ll\rlight$. This behaviour is expected as well inside the light-cylinder the influence of rotation and magnetospheric currents become negligible, the first being of order $(R/\rlight)^2$ and the second  of order $(R/\rlight)$. The FFE case produces higher curvature because of the stronger magnetic field sweep back due to the current flowing along field lines. For $R_1/\rlight=0.1$, typical values at this emission height are $\kappa_{\rm c} \, \rlight \approx 2.4$.

\begin{figure}[h]
	\centering
	\includegraphics[width=0.9\linewidth]{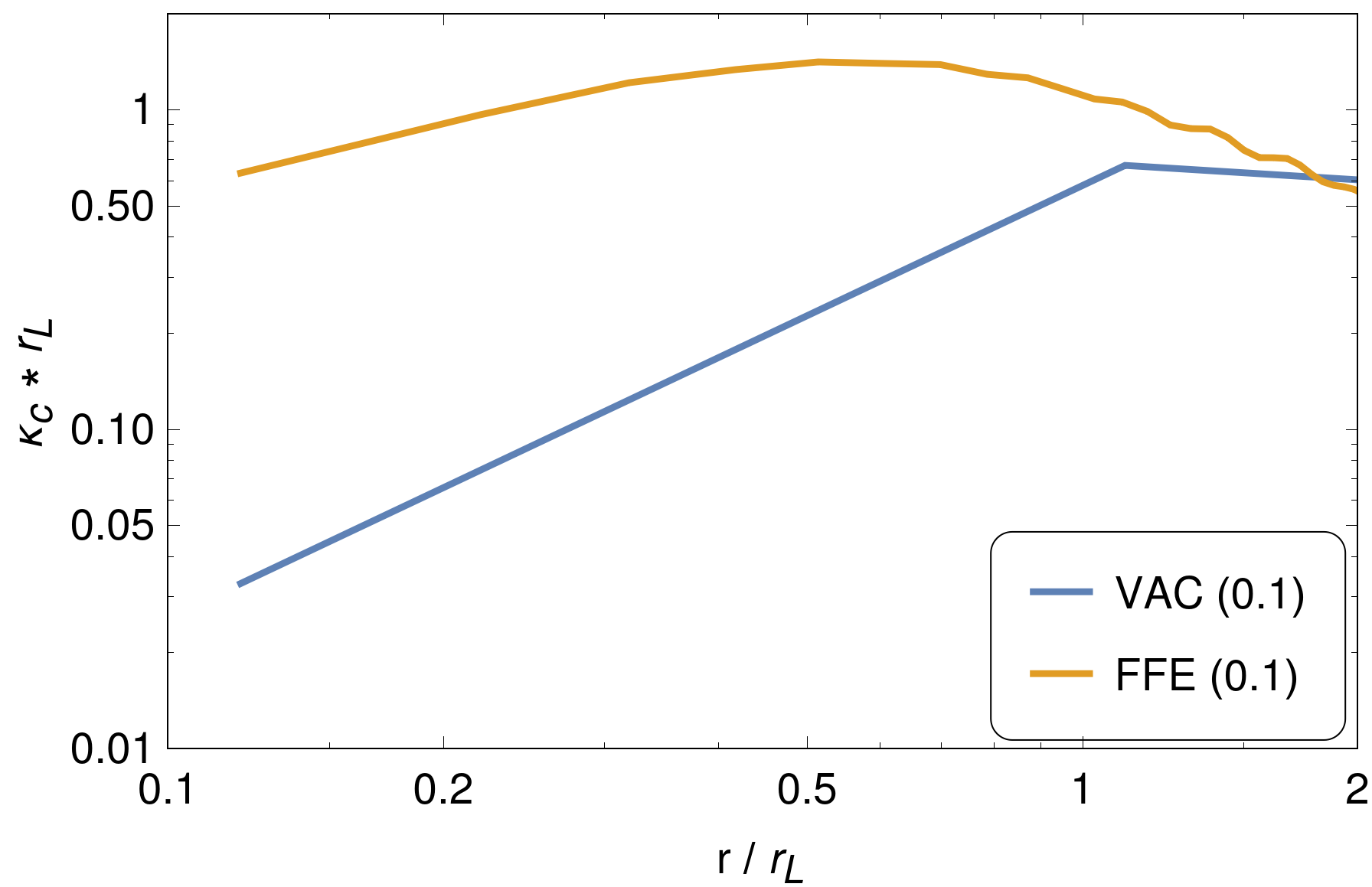}
	\caption{Curvature of the magnetic axis for the vacuum field (VAC) and the force-free magnetosphere (FFE). Values are shown for an obliquity of $\rchi=75\degr$.}
	\label{fig:courbure_vide_ffe}
\end{figure}

\begin{figure}[h]
	\centering
	\includegraphics[width=0.9\linewidth]{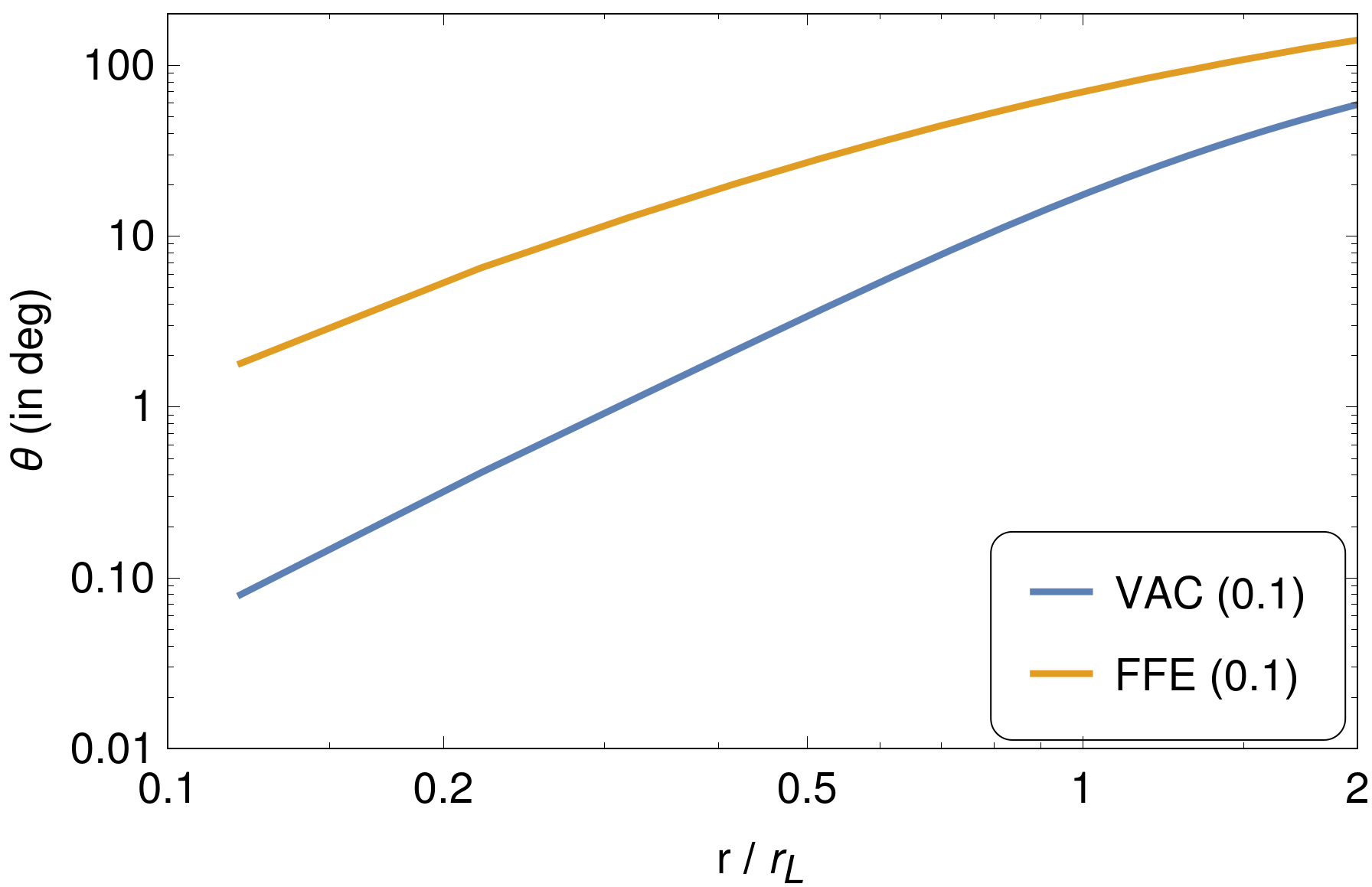}
	\caption{Angle between the local magnetic field at radius~$r$ for points on the central field line and the magnetic moment direction for the vacuum field (VAC) and the force-free solution (FFE). Values are shown for an obliquity of $\rchi=75\degr$.}
	\label{fig:angle_tx_vide_ffe}
\end{figure}

The behaviour of the curvature~$\kappa_c$ along the central magnetic field line is very different as shown in Figure~\ref{fig:courbure_vide_ffe}, in units of $1/\rlight$ for the vacuum Deutsch solution in blue (VAC) and for the force-free magnetosphere in orange (FFE). The curvature increases with distance to the star like $r^2$ for the vacuum and like $r^{1/2}$ for the FFE, as long as $r\ll\rlight$. This is in contradiction with the radius-to-frequency mapping because the high frequency photons are produced at higher altitude in this picture. We therefore anticipate an opposite variation of the radio pulse width between the core and the cone components as observed in some pulsars taken from the literature. Indeed \cite{posselt_thousand-pulsar-array_2021} found a sample of pulsars showing the opposite of the RFM prediction, that is a pulse width increase with frequency. Similar conclusions have been drawn by \cite{johnston_multifrequency_2008} who claimed that almost one-third of their sample disagrees with RFM. \cite{chen_frequency_2014} reported the same trend for about 20\% of 150 pulsars analysed, see also \cite{noutsos_pulsar_2015}. For a given frequency, the cone and core components emanate from different altitudes.
A last important point concerns the possible phase lag between radio pulses in different frequency band. Indeed, if emitted at different altitudes along the magnetic axis, because of the increase in bent field lines, we should related this lag to the difference in altitude as already mentioned by \cite{phillips_radio_1992}. To this end, we plot the angle between the local magnetic field at a radius~$r$ and the magnetic moment direction for the vacuum field (VAC) and the force-free solution (FFE) as shown in Figure~\ref{fig:angle_tx_vide_ffe}. The difference between VAC and FFE is significant, a factor 10 or more. Differences in emission altitudes will not only impact the width of the pulse profiles because of the increase of the radio cone beam opening angle, but also the phase of their centre due to the magnetic field sweep back.

\subsection{Broadband photon emission}
From a theoretical point of view, in the pair cascade region above the polar cap, current models predict two populations of pair plasma flows: a primary beam of electrons/positrons with very high Lorentz factor and accelerated in the vacuum gap potential drop to reach $\gamma_{\rm b} \approx 10^6$ and a secondary plasma of $e^\pm$ pairs produced by cascades due to magnetic photo-disintegration reaching Lorentz factors of $\gamma_{\rm p} \approx 10^2$ \citep{kazbegi_circular_1991, arendt_pair_2002, usov_two-stream_2002}. In the partially screened gap model of \cite{gil_drifting_2003} an ion outflow is allowed with $\gamma_{\rm ion}\approx 10^3$. Electrons and positrons do not follow exactly the same distribution functions because of the parallel electric field screening \citep{beskin_physics_1993}. The pair multiplicity factor can reach values up to $\kappa \approx 10^4-10^5$ \citep{timokhin_maximum_2019}. This pair plasma produces the observed radio emission, typically in the MHz-GHz band, through curvature radiation \citep{mitra_nature_2017}. 

On top of the primary and secondary beams, three different sites produce photons at different energies. Radio emission is a consequence of curvature radiation along open magnetic field lines close to the polar caps. The Lorentz factors requires to emit typically at a frequency of 1~GHZ at a height of $r/\rlight \approx 0.1$ where the curvature is about $\kappa_{\rm c} \, \rlight = \rlight / \rho_{\rm c} \approx 22-35$ are
\begin{equation}
	\gamma_{\rm radio} \approx 57 \, \left( \frac{\nu_{\rm radio}}{1~\textrm{GHz}} \right)^{1/3} \, \left( \frac{30 \, \rho_{\rm c}}{\rlight}\right)^{1/3}
\end{equation}
which corresponds to the secondary plasma flow as expected.

For the X-ray photons we could either invoke synchrotron radiation or curvature radiation. The particle Lorentz factor to produce typically a 5~keV photon must be in either case
\begin{subequations}
	\begin{align}
		\gamma_{\rm sync} & \approx \left( \frac{2}{3} \, \frac{\omega_{\rm sync}}{\omega_{\rm B} \, \sin\psi} \right)^{1/2} \approx 1358 \, \left( \frac{E_{\rm X}}{5~\textrm{keV}} \right)^{1/2} \, \left( \frac{B}{B_{\rm L}}\right)^{-1/2} \\
		\gamma_{\rm curv} & \approx \left( \frac{2}{3} \, \frac{\omega_{\rm curv} \, \rho_{\rm c}}{c}\right)^{1/3} \approx 3.5\times10^5 \, \left( \frac{E_{\rm X}}{5~\textrm{keV}} \right)^{1/3} \, \left( \frac{\rho_{\rm c}}{\rlight}\right)^{1/3}
	\end{align}
\end{subequations}
where $\psi$ is the particle pitch angle with respect to the magnetic field line. For the numerical value, we used the most favorable pitch angle of $\psi=90\degr$. However, synchrotron radiation is very unlikely because particles stay in their fundamental Landau level up to the light cylinder.

If the X-ray photons are produced in regions with similar curvature as radio photons, then the particle Lorentz factor ratio must be
\begin{equation}
	\frac{\gamma_{\rm X}}{\gamma_{\rm radio}} = \left( \frac{5~\textrm{keV}}{h\times1~\textrm{GHz}}\right)^{1/3} \approx 1065
\end{equation}
which corresponds to the primary beam with low Lorentz factor of about $10^{4-5}$. 
Therefore radio and non-thermal X-ray emission are produced by curvature radiation of the secondary plasma ($\gamma_{\rm b} \approx 100$) and the primary beam ($\gamma_{\rm p}\gtrsim10^5$) respectively. This is consistent with the one-dimensional particle distribution function of the out-flowing relativistic plasma along open magnetic field lines.
Moreover, above a height of $0.5\,\rlight$ the sharp decrease of the curvature~$\kappa_{\rm c}$ shifts the photon energy to a lower band well below 1~keV. In addition, the particle density number drops due to the divergent magnetic field geometry and the spherical expansion.

The reason why non-thermal X-rays are mostly produced along the separatrix is two fold. First, the curvature radiation power scales as the curvature squared $\kappa_{\rm c}^2$  and the particle charge squared $q^2$ such that
\begin{equation}
	\mathcal{P}_{\rm c} = \frac{q^2}{6\,\pi\,\varepsilon_0} \, \gamma^4 \, c \, \kappa_{\rm c}^2 .
\end{equation}
Because the curvature $\kappa_{\rm c}$ drastically decreases towards the centre of the polar cap, the associated curvature power also decreases, even faster that $\kappa_{\rm c}$. This leads to a hollow cone model reminiscent of the radio hollow cone model. Second, the particle density number~$n_e$ along the separatrix is high due to the electric current required to support the transition layer between the open field line region and the close field line region. Moreover, the current density decreases towards the centre for an inclined rotator and vanishes for an orthogonal rotator, see figure~8 and 9 of \cite{petri_spheroidal_2022}. Because the emissivity is proportional to the product $n_e\,\mathcal{P}_{\rm c}$, we expect the light-curve to be formed essentially in the separatrix region as postulated in our model. Adding some small resistivity to create a parallel accelerating electric field would only slightly change the value of the curvature~$\kappa_{\rm c}$ and not impacting the magnetic field within the magnetosphere. We next link the above discussion to observations from the third pulsar catalogue \citep{smith_third_2023}, extracting some statistical significance of the pulsar sub-populations.

\section{Statistic of radio, X-ray and $\gamma$-ray pulsars}
\label{sec:Statistics}

\begin{figure}[h]
	\centering
	\includegraphics[width=0.5\linewidth]{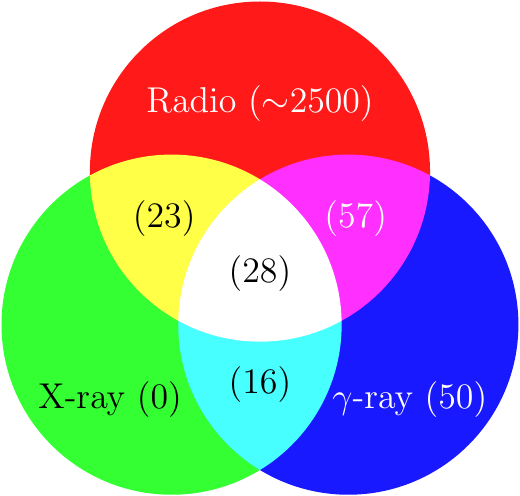}
	\caption{Venn diagram of the normal pulsar population showing their repartition in the different wavelengths, radio only in red, X-ray only in green and $\gamma$-ray only in blue. Others colours mean counts the pulsars detected at least in two of these wavelengths.
	\label{fig:radioxgammavendiagram}}
\end{figure}

\begin{figure}[h]
	\centering
	\includegraphics[width=0.5\linewidth]{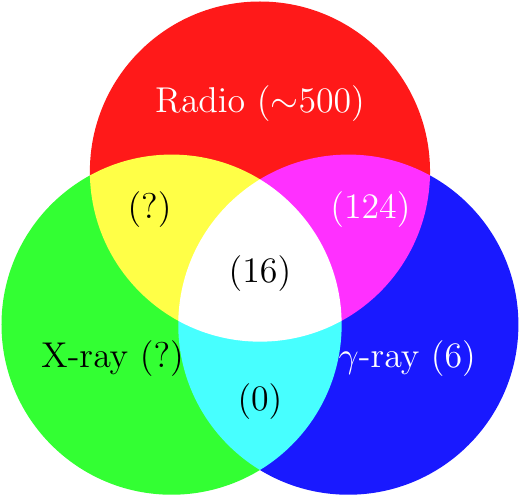}
	\caption{Same as fig.~\ref{fig:radioxgammavendiagram} but for millisecond pulsars. The question mark indicates that it was not possible to count this pulsars. No meaningful count for X-ray only or radio-loud X-ray pulsars were found, these are marked by a question mark (?).
	\label{fig:radioxgammamspvenndiagram}}
\end{figure}

In this section we explore the statistics of detected $\gamma$-ray pulsars seen also in radio and in X-ray. According to the latest update of the third pulsar catalogue available at \url{https://confluence.slac.stanford.edu/display/GLAMCOG/Public+List+of+LAT-Detected+Gamma-Ray+Pulsars}, 297~$\gamma$-ray pulsars are listed, see table~\ref{tab:gamma-ray_3PC}. Some of them a radio-quiet (q) or radio-loud (r), some are seen pulsating in X-ray (x), some are part of a binary system (b) and some are millisecond pulsars (m). The third pulsar catalogue of \cite{smith_third_2023} reports also the $\gamma$-ray time lag~$\delta$ and peak separations~$\Delta$ useful for our statistical study. We added some pulsars detected in non-thermal X-rays as reported by \cite{cotizelati_spectral_2020} and \cite{chang_observational_2023}. A graphical summary of young and millisecond pulsars is given as Ven diagrams respectively in fig.~\ref{fig:radioxgammavendiagram} and in fig.~\ref{fig:radioxgammamspvenndiagram}.
\begin{table}[h]
	\centering
\begin{tabular}{ccccc}
	\hline
	PSR & \multicolumn{2}{c}{normal} & \multicolumn{2}{c}{millisecond} \\
	\hline
	 & \multicolumn{2}{c}{151} & \multicolumn{2}{c}{146} \\
	 \hline
	 & r & q & mr & mq \\
	 & 85 & 66 & 140 & 6 \\
	 & \multicolumn{4}{c}{X-ray detection}  \\	
	\hline
	 & 51 & 16 & 16 & 0 \\
	\hline
\end{tabular}
	\caption{Statistic of 297~$\gamma$-ray pulsars as detected by Fermi/LAT. The last line corresponds to those detected in X-ray. The different letters stand for, r : radio-loud, q : radio-quiet, m : millisecond, x : X-ray. \label{tab:gamma-ray_3PC} }
\end{table}

\subsection{Radio-$\gamma$-ray population study}

\begin{figure}[h]
	\centering
	\includegraphics[width=0.9\linewidth]{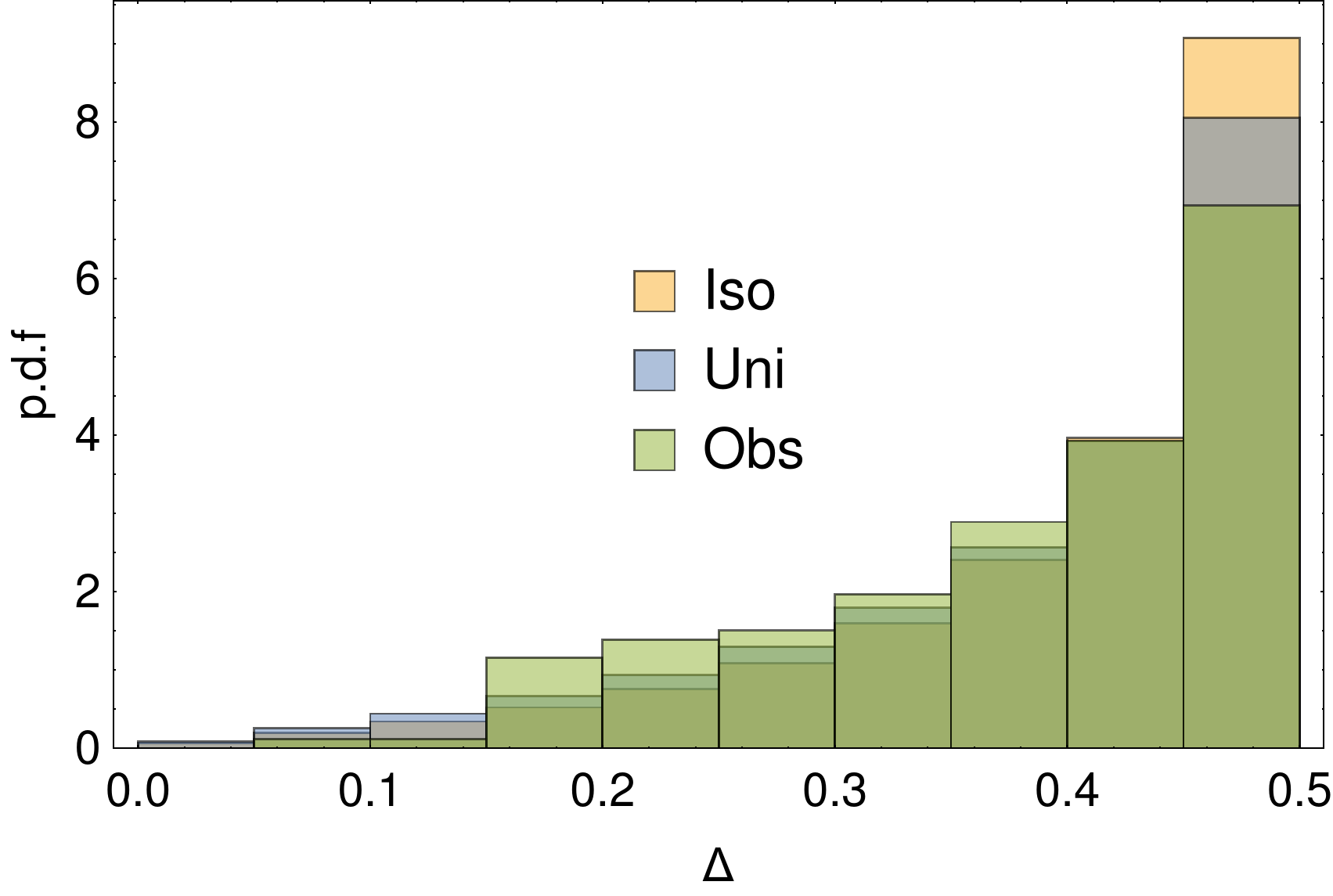}
	\caption{The $\gamma$-ray peak separation probability density function (p.d.f.) obtained from the observations (Obs) of 3PC in green versus the model prediction for an isotropic (Iso) obliquity distribution in orange and an uniform (Uni) distribution in blue.}
	\label{fig:delta2pcmodel}
\end{figure}

Our aim is to check whether the detected $\gamma$-ray pulsar distribution is compatible with an isotropic or uniform distribution of magnetic obliquity~$\rchi$ and line of sight inclination angle~$\zeta$ assuming a striped wind model for $\gamma$-rays and a polar cap model for the radio band. To this end, we performed Monte-Carlo simulations, throwing ten million random numbers~$Z$, uniformly distributed in the interval $[0,1]$ for each angle $\rchi \in [0,\pi]$ and $\zeta \in [0,\pi]$. The isotropic angle distribution depicted by the random variable~$\Theta$ is then deduced by the change of variable
\begin{equation}
	\Theta = \arccos (Z)
\end{equation}
where $\Theta \in [0,\pi]$ is meant for either $\rchi$ or $\zeta$. The probability density function of both angles therefore follow a $\sin$ law according to
\begin{equation}
	p(\Theta = \theta ) = \frac{\sin \theta}{2} \qquad ; \qquad \theta \in [0,\pi] .
\end{equation}
For an uniform distribution of obliquities we choose an angle from the distribution
\begin{equation}
	\Phi = \pi \, Z
\end{equation}
the angle $\zeta$ remaining isotropic. This distribution would mimic a tendency towards alignment of the pulsar obliquity. First we compare the distribution of the $\gamma$-ray peak separation~$\Delta$ given by \cite{smith_third_2023} to our simulated sample knowing that a simple relation exists between $\Delta$, $\rchi$ and $\zeta$, given by \cite{petri_unified_2011} as
\begin{equation}\label{eq:separation_pic}
	\cos(\pi\,\Delta) = |\cot \rchi \, \cot \zeta| .
\end{equation}
Our algorithm is therefore straightforward. We throw ten~million instances of $\rchi$ and independently ten~million instances of $\zeta$ from an isotropic or a uniform distribution in $\rchi$ and deduce $\Delta$ from eq.~\eqref{eq:separation_pic}. We restrict the separation to the interval $\Delta \in [0,0.5]$. If the observed $\Delta$ is larger than $0.5$ we compute $1-\Delta$ because $\Delta>0.5$ can be interpreted as $\Delta<0.5$ if the order of the $\gamma$-ray peaks is inverted. Applying this transform, the results are summarized in the histogram of figure~\ref{fig:delta2pcmodel} showing the probability density function (p.d.f.). The observations are depicted by green rectangles whereas the simulated samples by orange rectangles for an isotropic $\rchi$ distribution and blue rectangles for a uniform $\rchi$ distribution. The agreement is satisfactory although we overestimated slightly the probability of a separation close to half a period $\Delta \approx 0.5$ in both cases, although a uniform distribution gives slightly better results.

\begin{figure}[h]
	\centering
	\includegraphics[width=0.9\linewidth]{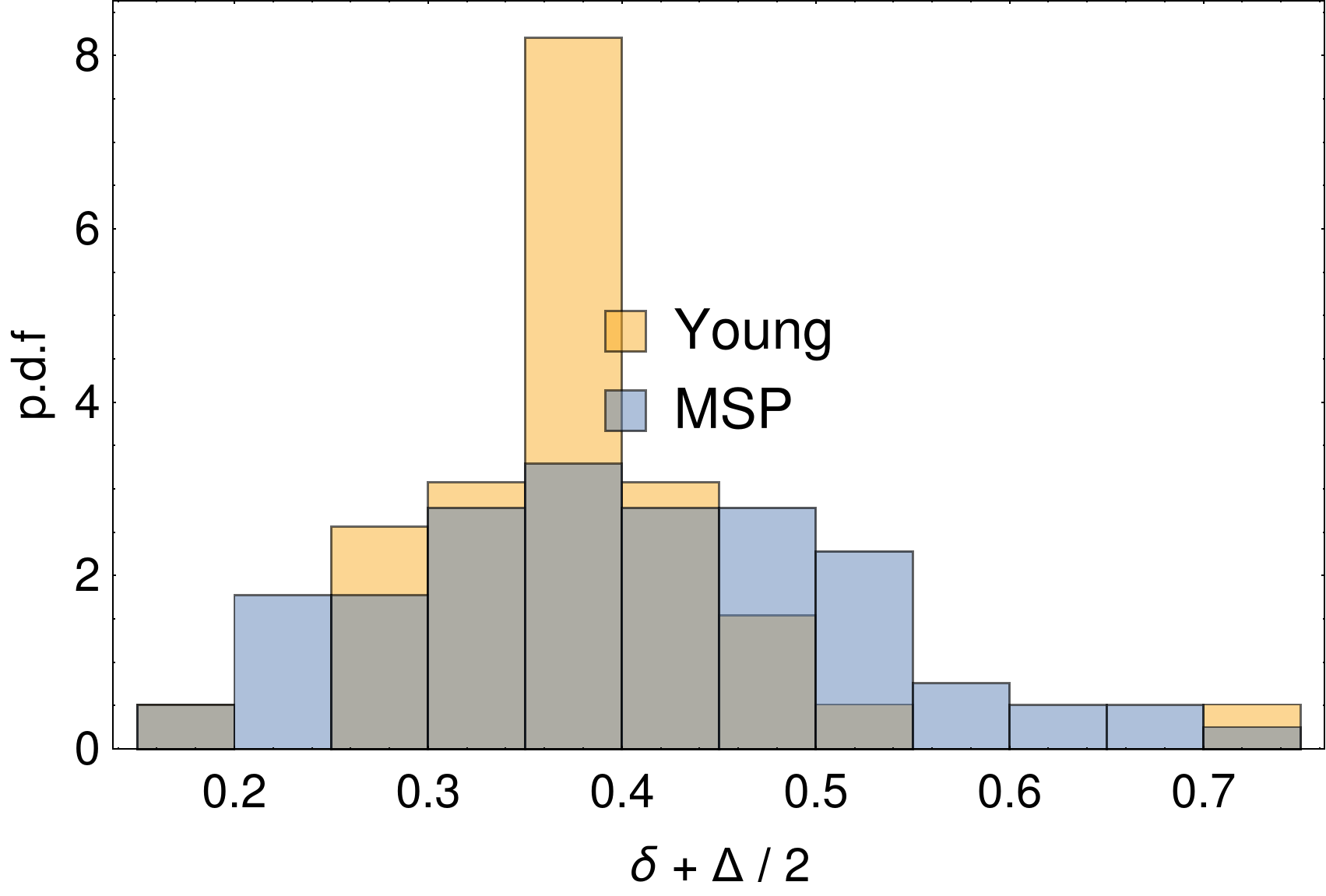}
	\caption{The relation between time lag and peak separation as given by eq~\eqref{eq:radiolag_peak_separation} shown for the sample of young and millisecond pulsars from 3PC.}
	\label{fig:radiolag_peak_separation}
\end{figure}

In a previous work \cite{petri_young_2021} showed that the $\gamma$-ray lag~$\delta$ is related to the peak separation~$\Delta$ by an approximate relation such as
\begin{equation}\label{eq:radiolag_peak_separation}
	\delta + \frac{\Delta}{2} \approx 0.46-0.5 .
\end{equation}
For high line of sight inclinations, it becomes closer to 0.46. We checked if this relation is satisfied in the 3PC pulsar catalogue, for young and millisecond pulsars independently. The results are shown in the histograms of figure~\ref{fig:radiolag_peak_separation}. For the young pulsar population, we observe a strong peak in the range $0.25-0.5$, thus a time lag about 0.1 less in phase than the one predicted by the force-free model. There is a systematic effect that is missed in our description. This shift has already been noticed by \cite{petri_unified_2011}. The millisecond pulsars show a similar trend but less peaked around $0.25-0.5$ and values more spread between 0.1 and 0.8. We do not expect millisecond pulsars do follow sharply eq.~\eqref{eq:radiolag_peak_separation} because this formula assumes a dipolar field in the radio emission region which is not the case for this population. 
The NICER collaboration found hot spots far from a geometry for a centred dipole. Off-centred or even multipole components are favoured as found by the studies of \cite{riley_nicer_2019} and \cite{salmi_radius_2022}. Moreover \cite{rigoselli_new_2018} found absorption lines in the spectrum of PSR B1133+16, interpreted as proton cyclotron features, thus implying multipolar components at the stellar surface. Other hints are given by \cite{schwope_phase-resolved_2022} for J0659+1414 showing a complex variation in surface temperature suggesting also the presence of multipolar field.

\begin{figure}[h]
	\centering
	\includegraphics[width=\linewidth]{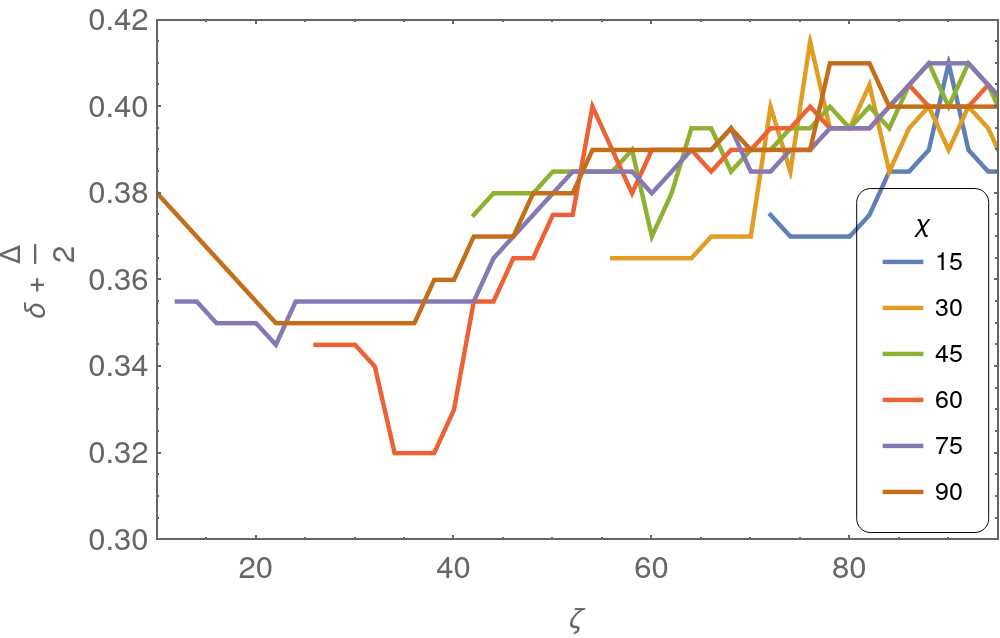}
	\caption{The quantity $\delta + \Delta/2$ for the pulsar wind particle velocity oriented in the forward direction with an angle of $45\degr$ with respect to the radial direction.}
	\label{fig:fitdelairi0}
\end{figure}

This discrepancy can be explained by the fact that the high-energy radiation is not directed radially forward but is subject to a beaming in the direction of corotation that is not included in our striped wind model. Let us show how this proceeds by assuming particles moving at the speed of light along magnetic field lines in the equatorial plane for an orthogonal rotator. The particle velocity is $\vec{\beta} = \vec{\Omega} \wedge \vec{r} + \alpha \, \vec{t}$ where $\vec{t}$ is the unit vector tangent to the local field line. The angle between the radial direction and $\vec{t}$ is denoted by $\theta$ thus $\cos\theta = \er \cdot \vec{t}$. Introducing a local planar Cartesian coordinate system $(O,x,y)$ with the origin~$O$ at the light-cylinder where emission starts, the speed is $\beta^2 = \alpha^2 - 2\,\alpha\,\sin\theta +1$. For ultra-relativistic speeds $\beta \lesssim 1$ and $\alpha=0$ or $\alpha=2\,\sin\theta$. The first solution must be rejected because it corresponds to a straight line but field line as swept back. Thus the second solution applies. The angle~$\phi$ between the particle velocity and the radial direction (the $x$-axis) becomes $\tan\phi = v_y/v_x = \cot (2\,\theta)$. The solution therefore is simply 
\begin{equation}\label{eq:phi_theta_gamma}
\phi = \frac{\pi}{2} - 2\,\theta .
\end{equation}
The angle $\phi$ introduces an additional time lag between radio and $\gamma$-ray of $\phi/(2\pi)$ in phase. Observations require this to be around 0.1 thus $\phi \approx \pi/5$ and the magnetic field sweep-back angle is $\theta \approx 3\pi/20 = 27 \degr$. We conclude that the magnetic field line must bent counter-clockwise at 27\degr to account for this additional shift. This value is compatible with a force-free neutron star magnetosphere. Variation in the bending leads to a spread in the time lag. 
In order to check this idea in the general case, we computed new $\gamma$-ray light-curves by orienting the velocity vector in the forward direction with an inclination of $45\degr$ with respect to the radial direction. The quantity $\delta + \Delta/2$ is shown in fig.~\ref{fig:fitdelairi0} where it ranges from $0.32$ to $0.42$ which corresponds to the peak in the histogram of fig.~\ref{fig:radiolag_peak_separation}. With a non radial flow the agreement between observations and our model improves. The fact that the particle velocity right outside the light-cylinder must be non radial was already noticed by \cite{contopoulos_hybrid_2020} who found a decreasing azimuthal component and an increasing radial component.

Next let us investigate the proportion of radio-loud and radio-quiet $\gamma$-ray pulsars as well as $\gamma$-quiet pulsars. 
In order to decide whether a pulsar will be seen in $\gamma$-rays or not, we compute the quantity $G = \rchi - |\zeta - \pi/2|$, from eq.\eqref{eq:zeta_gamma_visible}. A positive value~$G$ means that the pulsar is visible in $\gamma$-rays. The figure~\ref{fig:visibilite_gamma} shows the probability density function of this quantity~$G$. By construction, the total area of this histogram equals unity and the area delimited by $G>0$ corresponds to the probability of detecting a $\gamma$-ray pulsar. We find that this area is equal to 0.89 therefore 89\% of the pulsar population should be seen as $\gamma$-ray pulsars.

\begin{figure}[h]
	\centering
	\includegraphics[width=\linewidth]{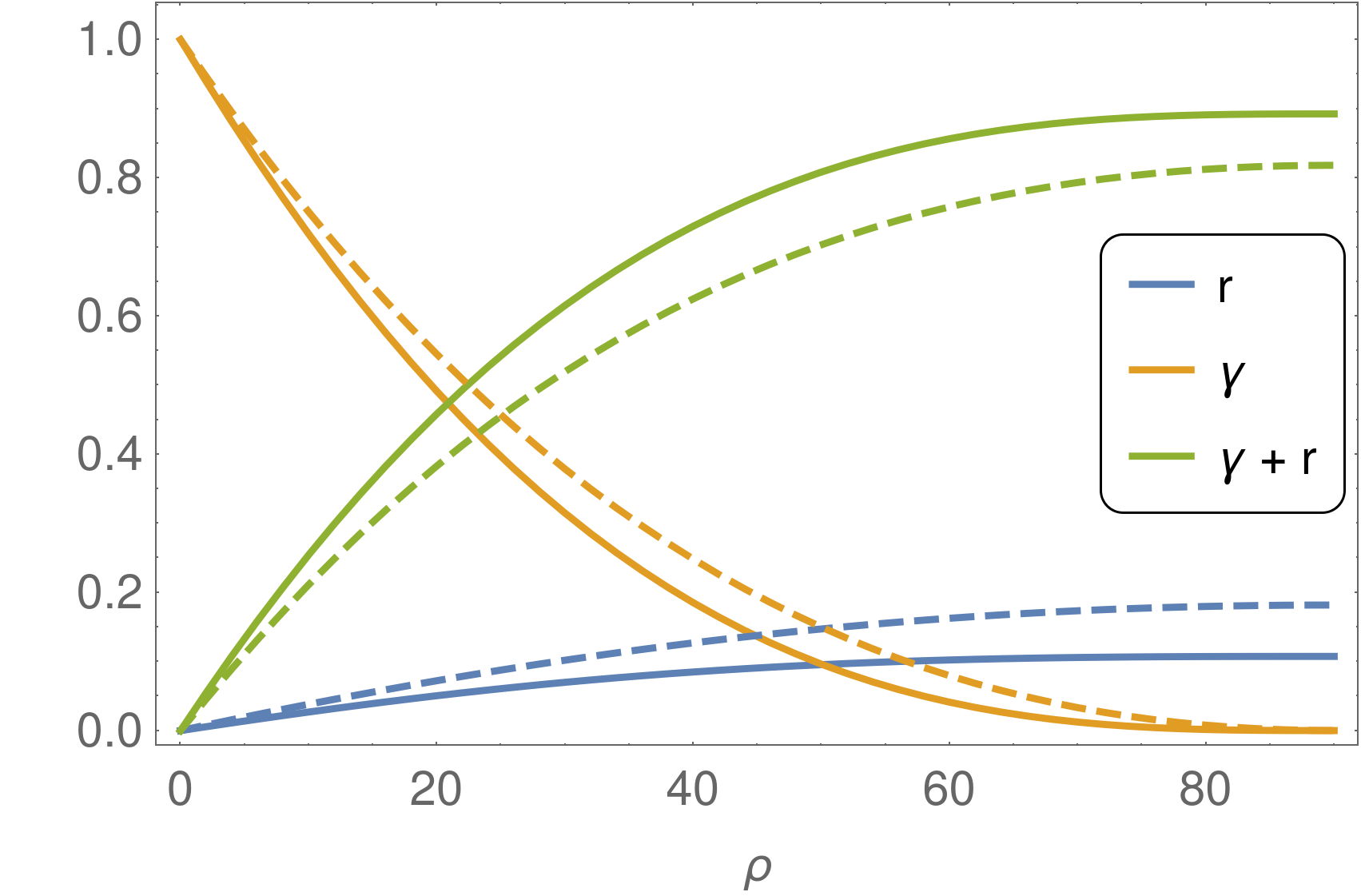}
	\caption{Fraction of radio-only pulsars "r", $\gamma$-only pulsars "$\gamma$" and radio-loud $\gamma$-ray pulsars "r+$\gamma$" as a function of the radio beam cone half opening angle $\rho$. The solid line represents the results for an isotropic~$\rchi$ whereas the dashed line for a uniform $\rchi$.}
	\label{fig:evolution_fraction_rhocone_1}
\end{figure}

\begin{figure}[h]
	\centering
	\includegraphics[width=\linewidth]{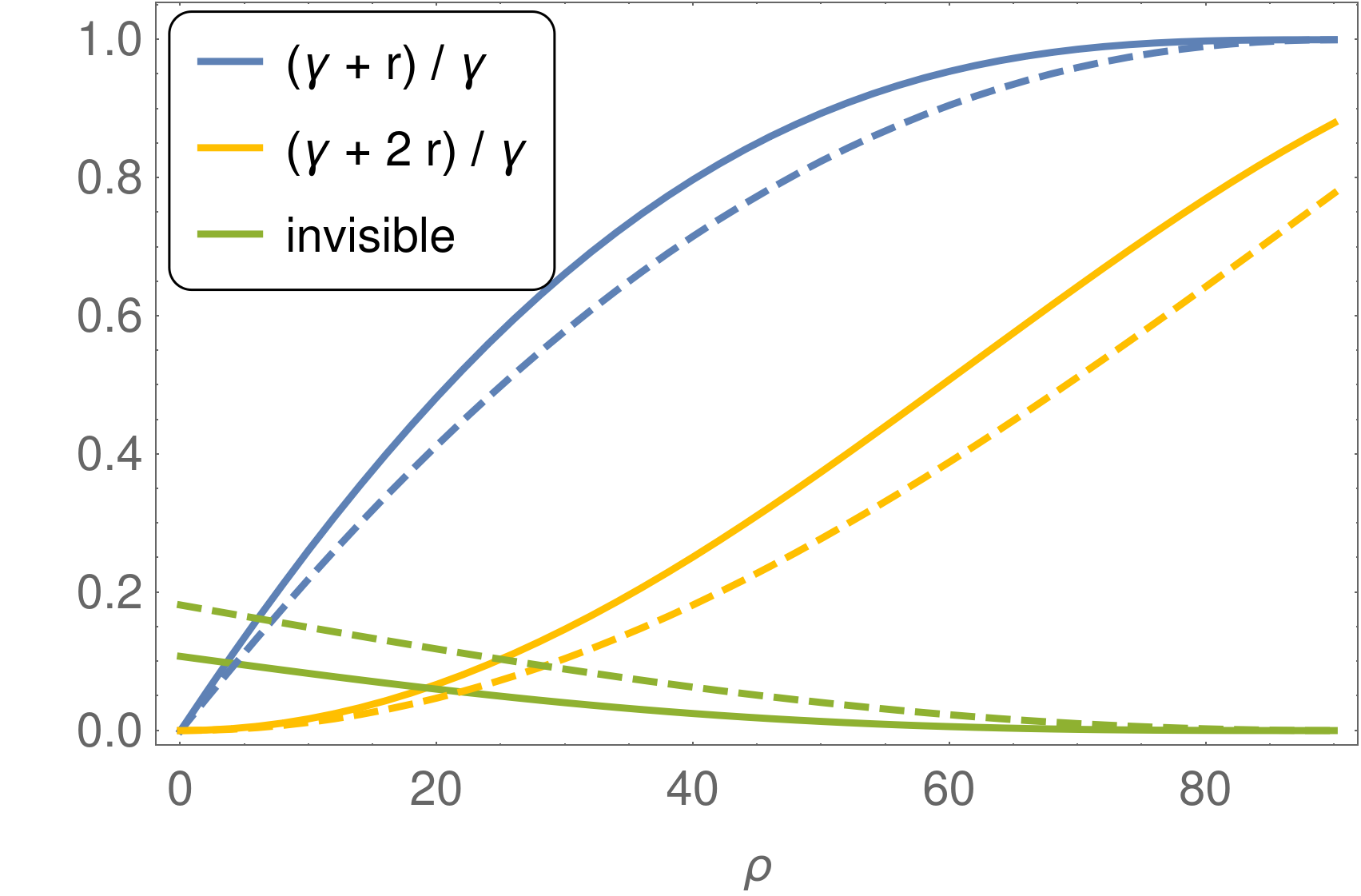}
	\caption{Fraction of radio-loud $\gamma$-ray pulsars with one peak $(\gamma+r)/\gamma$ and with two peaks $(\gamma+2r)/\gamma$ and fraction of invisible pulsars (not detected neither in r nor in $\gamma$). Solid line for isotropic $\rchi$ and dashed line for uniform $\rchi$.}
	\label{fig:evolution_fraction_rhocone_2}
\end{figure}

Fig.~\ref{fig:evolution_fraction_rhocone_1} shows the fraction of radio-only pulsars, $\gamma$-only pulsars and radio-loud $\gamma$-ray pulsars as a function of the radio beam cone half opening angle $\rho$. The solid line represents the results for an isotropic~$\rchi$ whereas the dashed line for a uniform $\rchi$. 
Fig.~\ref{fig:evolution_fraction_rhocone_2} shows the evolution of fraction of radio-loud $\gamma$-ray pulsars with one peak, two peaks and invisible pulsars (not detected neither in radio nor in $\gamma$) as a function of $\rho$. For a beam angle of $\rho \approx 25\degr$, half  of $\gamma$-ray pulsars are detected in radio.
Table~\ref{tab:pulsar_populations} summarized the simulated pulsar population for isotropic and uniform obliquity distributions in the different wavelengths.
\begin{table}[h]
	\centering
	\begin{tabular}{crr}
		\hline
		Wavelength & Isotropic & Uniform \\
		\hline
		invisible & 477.196 & 1.002.213 \\
		radio     & 594.977 & 813.714 \\
		$\gamma$-ray & 8.927.827 & 8.184.073 \\
		\hline
	\end{tabular}
	\caption{Summary of simulated pulsar population for isotropic and uniform obliquity distributions. The total number of pulsars in each column is $10^7$ and the radio beam opening angle $\rho=26\degr$. \label{tab:pulsar_populations}}
\end{table}
New $\gamma$-ray pulsar are still discovered in unassociated Fermi-LAT sources by recent surveys, like for instance the Transients and Pulsars with MeerKAT (TRAPUM) Large Survey Project \citep{clark_trapum_2023} or the Einstein@Home blind search survey \citep{wu_einsteinhome_2018}. There exists also several thousands of unidentified $\gamma$-ray sources in the Fermi catalogue. A significant fraction could be pulsars and the statistics discussed in this work could evolve in the future.

A similar study can be conducted for the radio-loudness. The relevant quantity are $N = |\zeta - \rchi|$ and $S = |\zeta + \rchi - \pi|$ according to eq.~\eqref{eq:zeta_radio_visible_nord} and eq.~\eqref{eq:zeta_radio_visible_sud}. Actually the condition $N \leq \rho$ is sufficient because both probability density functions are almost identical. Figure~\ref{fig:visibilite_radio} shows the probability density function of $N$ only as this is sufficient to decide whether a pulsar will be seen in radio or not. To complete we need to fix the radio beam cone opening angle $\rho$. The beam size depends on the pulsar period and on the emission height. For young pulsars, this height is constraint to lie around $h \approx 0.03-0.1 \, \rlight$ according to \cite{mitra_nature_2017} and to compiled data from \cite{weltevrede_profile_2008}.

The period distribution of the millisecond and young pulsar show a bi-modal shape well separated and sharply peaked around a mean value. The mean and median values are summarized in table~\ref{tab:pulsar_period_mean}. A typical young pulsar period is $P=150$~ms. The cone opening angle is therefore of the order
\begin{equation}\label{eq:cone_angle}
	\rho \approx \frac{3}{2} \, \sqrt{\frac{h}{\rlight}} 
	\approx (19-27) \deg.
\end{equation}
From table~\ref{tab:gamma-ray_3PC}, we conclude that $85/151 \approx 56$\% of the young $\gamma$-ray pulsars are detected in radio. To get this percentage requires $\rho \approx 24-28 \degr$. This would correspond to an average emission height of 8-11\% of $\rlight$.
For the MSP, $140/146 \approx 95$\% are detected as radio pulsars, thus almost all the population of MSP. This is due to the fact that the radio beam cone is much wider compared to young pulsars. This percentage requires a cone opening angle of $\rho \approx 108 \degr$ associated to an average emission height equal to a significant fraction of $\rlight$. In this case, the magnetic field has significant multipolar components and the expression for the cone width fails. We cannot infer any height for the radio emission, except being produced within the light cylinder, at a significant fraction of $\rlight$.
\begin{table}
	\centering
	\begin{tabular}{ccc}
		\hline
		& Mean (ms) & Median (ms) \\
		\hline
		MSP & 3.3 & 2.9 \\
		Young & 186 & 146 \\
		\hline
	\end{tabular}
\caption{Mean and median period of MSP and young pulsars in the Fermi catalogue. Values are given in ms.\label{tab:pulsar_period_mean}}
\end{table}

\begin{figure}
	\centering
	\includegraphics[width=\linewidth]{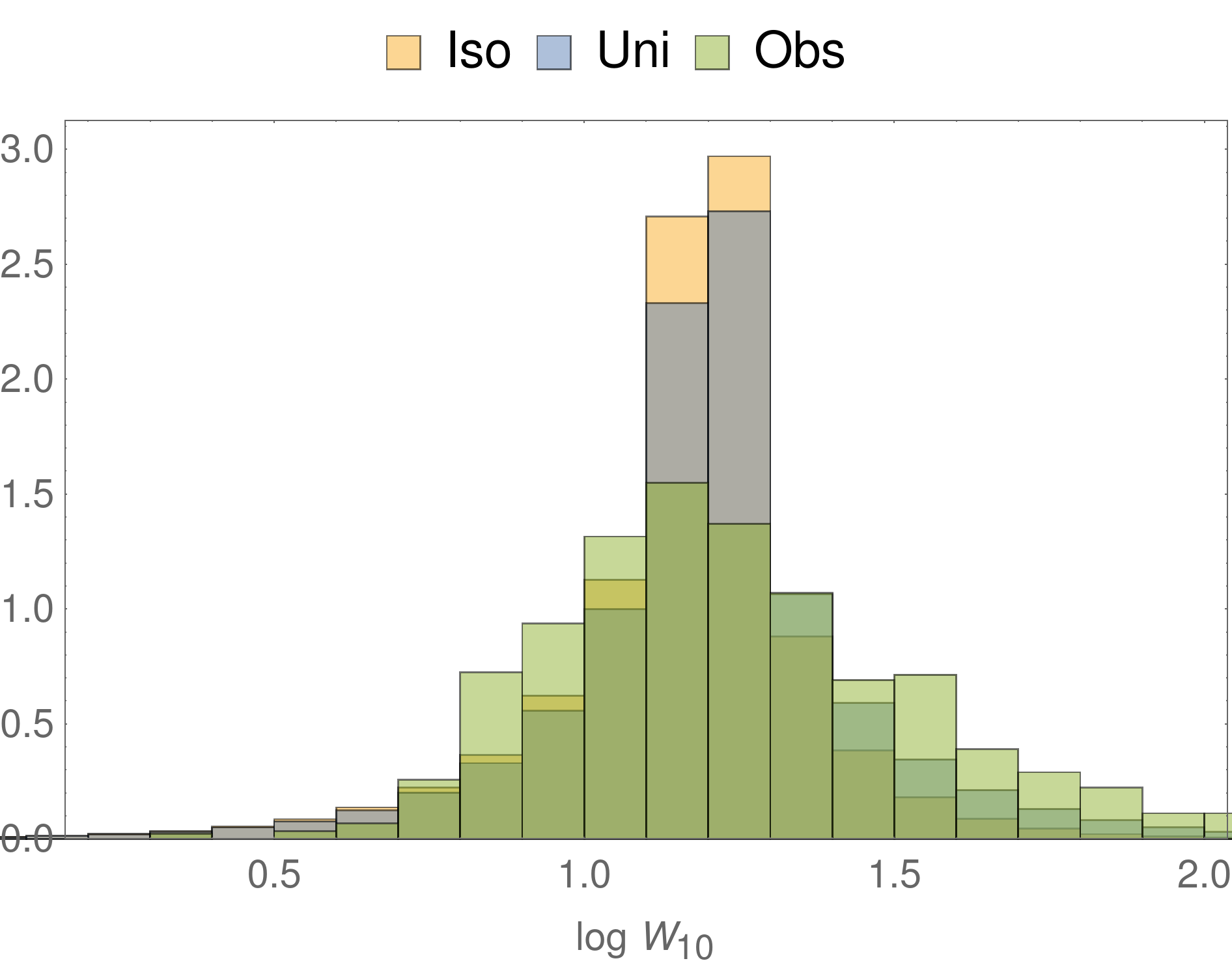}
	\caption{Pulse width $W_{10}$ at 10\% maximal height for an isotropic distribution of magnetic obliquity~$\rchi$, denoted "Iso", in orange, and a uniform distribution denoted "Uni", in blue, compared to the observations denoted by "Obs", in green.
	\label{fig:largeur_pulse}}
\end{figure}

Pulse width comparisons is another important check for the radio modelling. We use the data from \cite{posselt_thousand-pulsar-array_2021}. Fig~\ref{fig:largeur_pulse} shows the results of our model for an isotropic distribution of magnetic obliquity~$\rchi$, denoted by "Iso", in orange, and a uniform distribution denoted by "Uni", in blue, compared to the observations denoted by "Obs", in green. For a better match between our model and the observations, we took a radio emission cone opening angle of $\rho_{\rm em}\approx 16\degr$ which is about~2/3 of the radio cone beam width $\rho\approx 24\degr$.
We emphasize that these values are average quantities and that obviously all pulsars do not possess the same radio beam geometry, some being larger and some smaller than $\rho\approx 16\degr$. This scattering leads to a larger spread in the pulse width histogram shown in fig~\ref{fig:largeur_pulse}. We advise the reader to consult for instance the work of \cite{dirson_galactic_2022} for such refinement that is however out of the scope of the present paper.

\subsection{X-ray population study}

The data about thermal and non-thermal X-ray emission from pulsars are more scarce. No detailed catalogue exists such as in radio or $\gamma$-rays. It is therefore difficult to get a precise idea of the population of X-ray pulsars especially those showing pulsation in non-thermal X-rays. However, compiling data from \cite{chang_observational_2023} and \cite{cotizelati_spectral_2020} gives some hints about this particular population of pulsars. It is expected that more $\gamma$-ray sources and pulsar candidates will be associated to X-ray sources as revealed recently by \cite{mayer_searching_2024} with the SRG/eROSITA satellite. The census presented here serves therefore at most as a lower bound for the counting of pulsars shining in non-thermal X-rays. Orders of magnitude for the class of young and millisecond pulsars have been presented in fig.~\ref{fig:radioxgammavendiagram} and fig.~\ref{fig:radioxgammamspvenndiagram}.

To a first approximation, the population of X-ray pulsars follows the same trend as the radio pulsars because the emission geometries built onto a magnetic dipole, if well inside the light-cylinder, are very similar. The highest emission altitude determines the fraction of the sky illuminated by X-ray and therefore the fraction of pulsars detected in X-ray in addition to radio and/or $\gamma$-rays. Consequently, we can use the same fig.~\ref{fig:evolution_fraction_rhocone_1} and \ref{fig:evolution_fraction_rhocone_2} to deduce the expected number of detections in pulsed X-ray. The opening angle of the X-ray beam is much wider than the radio beam, leading to a large fraction of pulsars seen simultaneously in X-ray and $\gamma$-rays. But this hypothesis cannot be tested with high confidence because of the lack of an exhaustive census of X-ray pulsations.

Millisecond pulsars are difficult to fit into the picture of a dipolar field structure in the emission region. We therefore limit the X-ray expectations to the young non recycled pulsar population. The number of X-ray visible $\gamma$-ray pulsars is 44 for a total number of 151~$\gamma$-ray pulsars thus a fraction of $44/151 \approx 0.29 = 29$\% of X-ray loud $\gamma$-ray pulsars. This would correspond to an X-ray beam opening angle of $\rho_{\rm x} \approx 15 \degr$ and a low emission height along the separatrix, close to the radio emission site. We suspect that many pulsars should be detected in X-ray but the current low statistic does not allow us to put reliable constrain on the non thermal X-ray emission so far.

In order to test reliably the X-ray emission model in the current state, a different approach must be undertaken. One possibility is to resort to the fitting of individual pulsar X-ray light-curves, using geometrical constraints obtained from radio and $\gamma$-ray profile fitting as done by \cite{benli_constraining_2021} and \cite{petri_young_2021}. Young pulsars are the best targets because of the dominant dipole field in the emission regions. Good candidates are therefore $\gamma$-ray pulsars from the 3PC seen simultaneously pulsating in radio and X-ray. However, such a study goes far beyond the scope of this paper intended only to study pulsars as a population not individually.

\section{Conclusions}
\label{sec:Conclusions}

Radio as well as $\gamma$-ray emission regions are well constrained to be located above the polar caps for the former and in the current sheet of the striped wind for the latter. Nevertheless, the origin of the non-thermal X-ray radiation is less well established and only very few if no reliable constrained have been settled with high confidence. The comprehensive set of multi-wavelength pulse profiles computed in this work serves as an efficient and powerful tool to investigate and localise the X-ray photon production site by fitting jointly the radio pulse profile, the $\gamma$-ray light-curve and the non-thermal X-rays. As an application of the present work, in a subsequent paper, we will apply this fitting technique to several radio-loud X- and $\gamma$-ray pulsars and show how to reliably localize the non-thermal X-ray like for instance for PSR~J2229+6114, a pulsar bright enough in X-rays. Moreover based on the third pulsar catalogue, we were able to estimate an average radio emission altitude. We showed that the pulsar sub-populations of young and millisecond pulsars is compatible with an isotropic distribution of magnetic obliquity and line of sight inclinations angles although an evolution towards alignment is not excluded by our study.

This work is entirely based on geometrical effects to compute the light curves. No mention has been made about the energetics and dynamics of the charged particles evolving in this electromagnetic field, only some guesses are provided. As the force-free model cannot accelerate particles, it is difficult to estimated the Lorentz factor and the radiation energy output of this model. A good compromise between the ideal force-free flow and the fully kinetic description of the magnetosphere would be to investigate the resistive regime, allowing for a substantial parallel electric field able to accelerate particle to ultra-relativistic energies. The multi-wavelength light-curves would then depend on a new independent parameter depicting a kind of resistivity if an Ohm's law is employed. This is left as a future work.

\begin{acknowledgements}
I am grateful to the referee for helpful comments and suggestions that improved the quality of the paper. I would like to thank David Smith for stimulating discussions. This work has been supported by the CEFIPRA grant IFC/F5904-B/2018 and ANR-20-CE31-0010.
\end{acknowledgements}


\appendix

\section{Multi-wavelength atlas}

\begin{figure}[h]
	\centering
	\includegraphics[width=\linewidth]{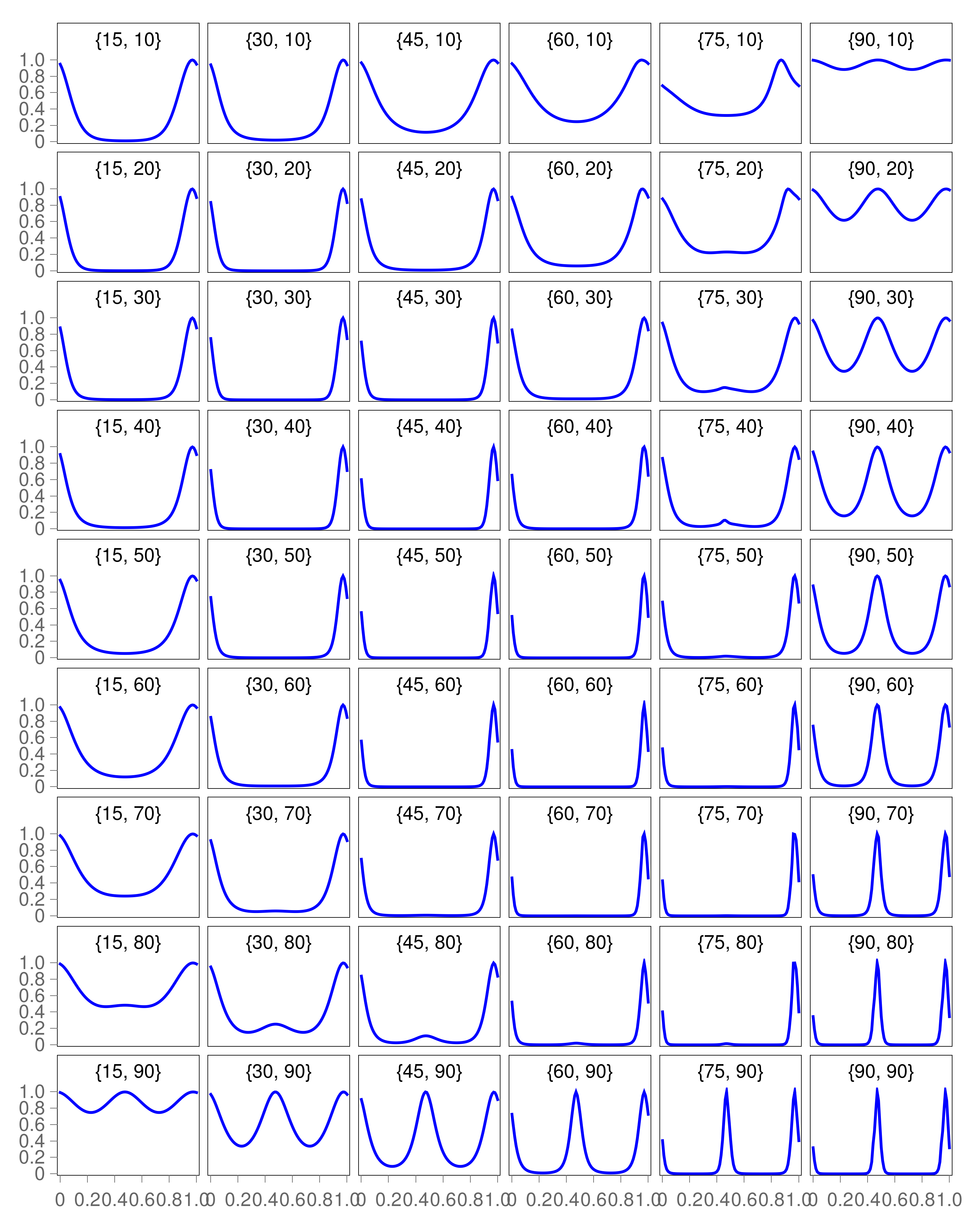}
\caption{Atlas of radio pulse profile for photons produced in the range $r/\rlight\in[0.1,0.11]$. The geometry is shown in the inset as brackets $\{\rchi, \zeta\}$. 
\label{fig:atlas_radio_r0.2}}
\end{figure}

\begin{figure}[h]
	\centering
	\includegraphics[width=\linewidth]{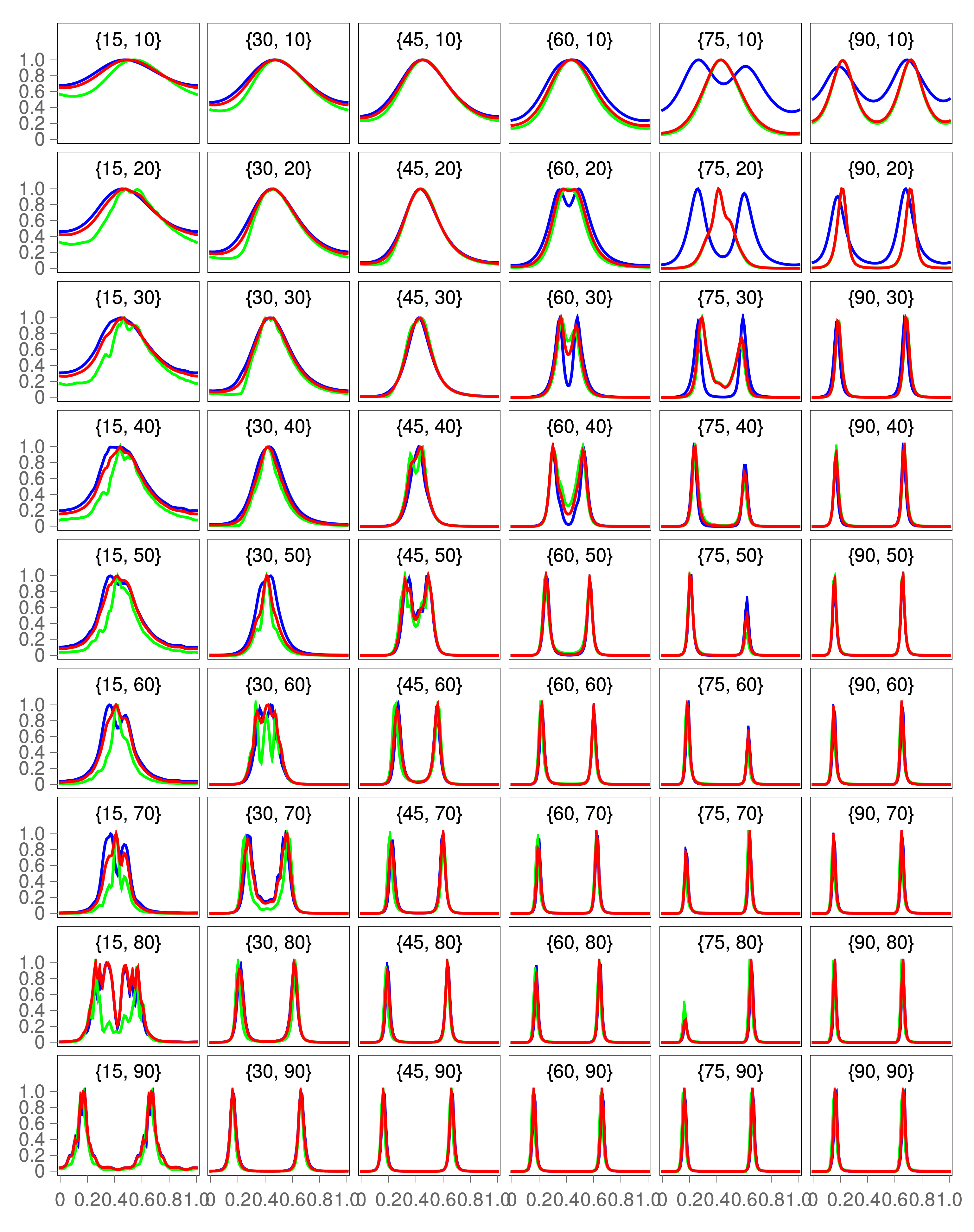}
	\caption{Atlas of $\gamma$-ray light-curves when emission emanates from the range $r/\rlight\in[1,2]$ in blue, in the range $r/\rlight\in[2,3]$ in green and the total $r/\rlight\in[1,3]$ in red. The geometry is shown in the inset as brackets $\{\rchi, \zeta\}$.}
	\label{fig:atlas_gamma_r1}
\end{figure}

\begin{figure}[h]
	\centering
	\includegraphics[width=\linewidth]{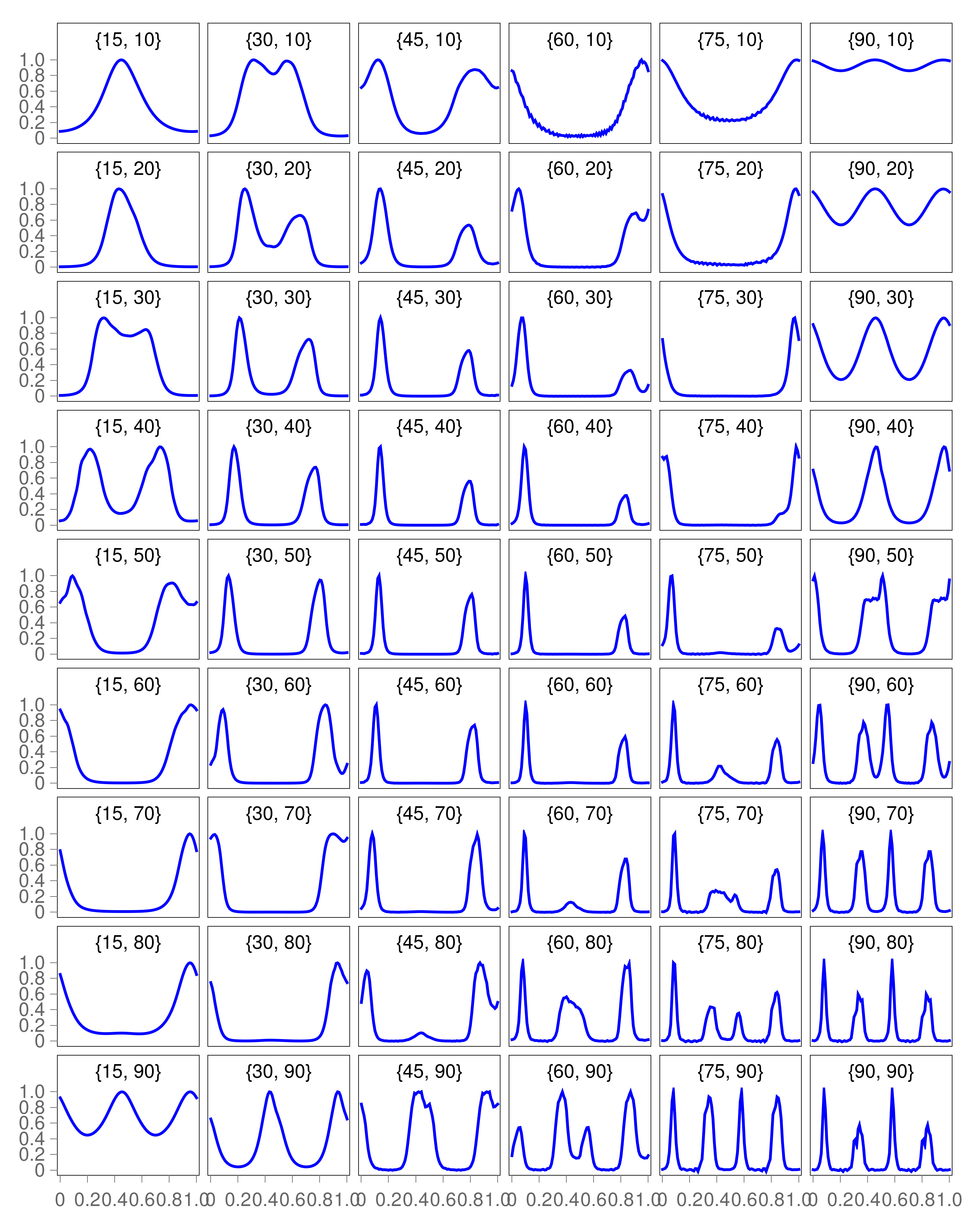}
	\caption{Atlas of X-ray light-curves light-curves when emission emanates from the range $r/\rlight\in[0.1,0.2]$. The geometry is shown in the inset as brackets $\{\rchi, \zeta\}$.}
	\label{fig:atlas_X_r0.2}
\end{figure}

\begin{figure}[h]
	\centering
	\includegraphics[width=\linewidth]{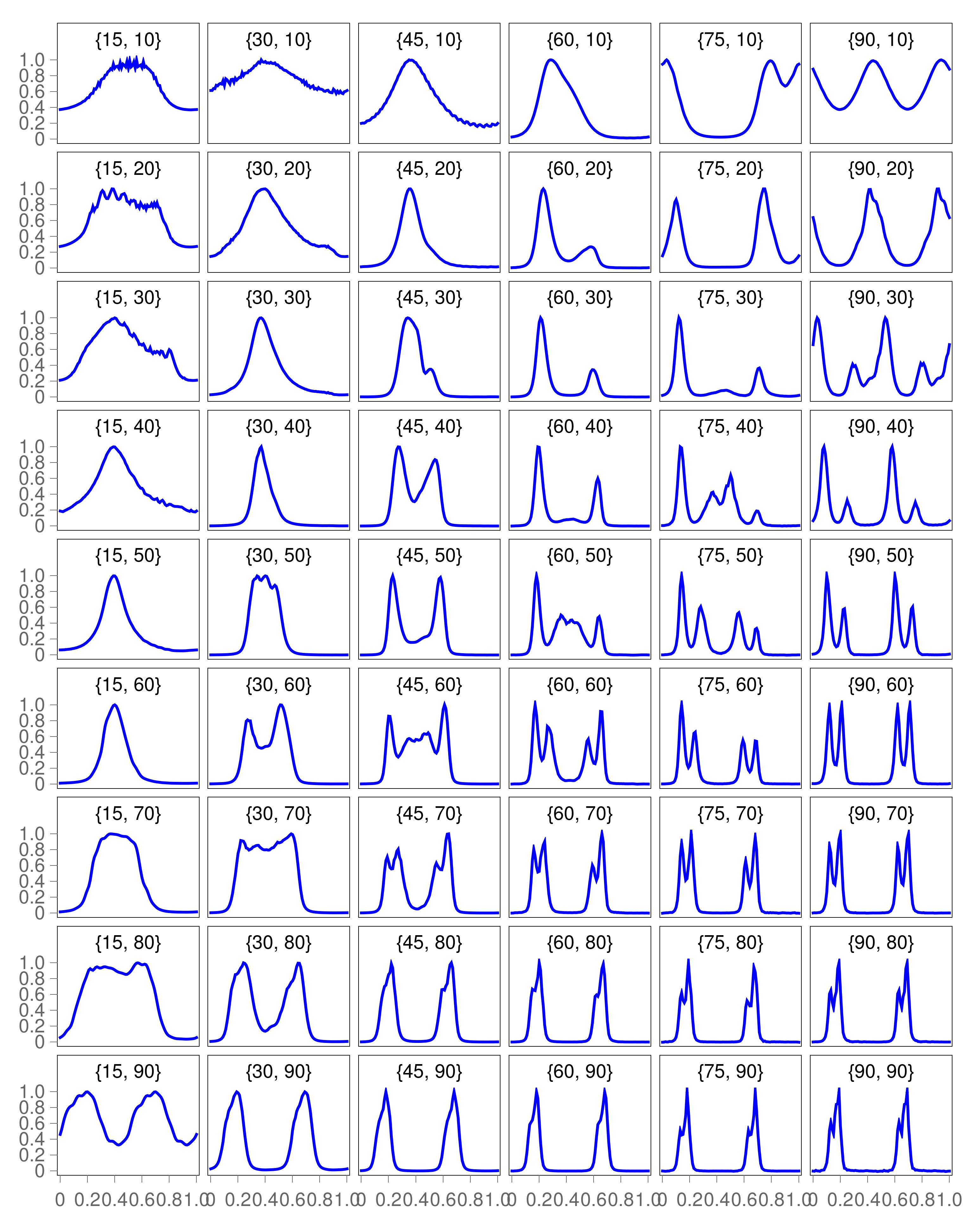}
	\caption{Atlas of X-ray light-curves light-curves when emission emanates from the range $r/\rlight\in[0.4,0.5]$. The geometry is shown in the inset as brackets $\{\rchi, \zeta\}$.}
	\label{fig:atlas_X_r0.5}
\end{figure}

\begin{figure*}
	\centering
	\includegraphics[width=\linewidth]{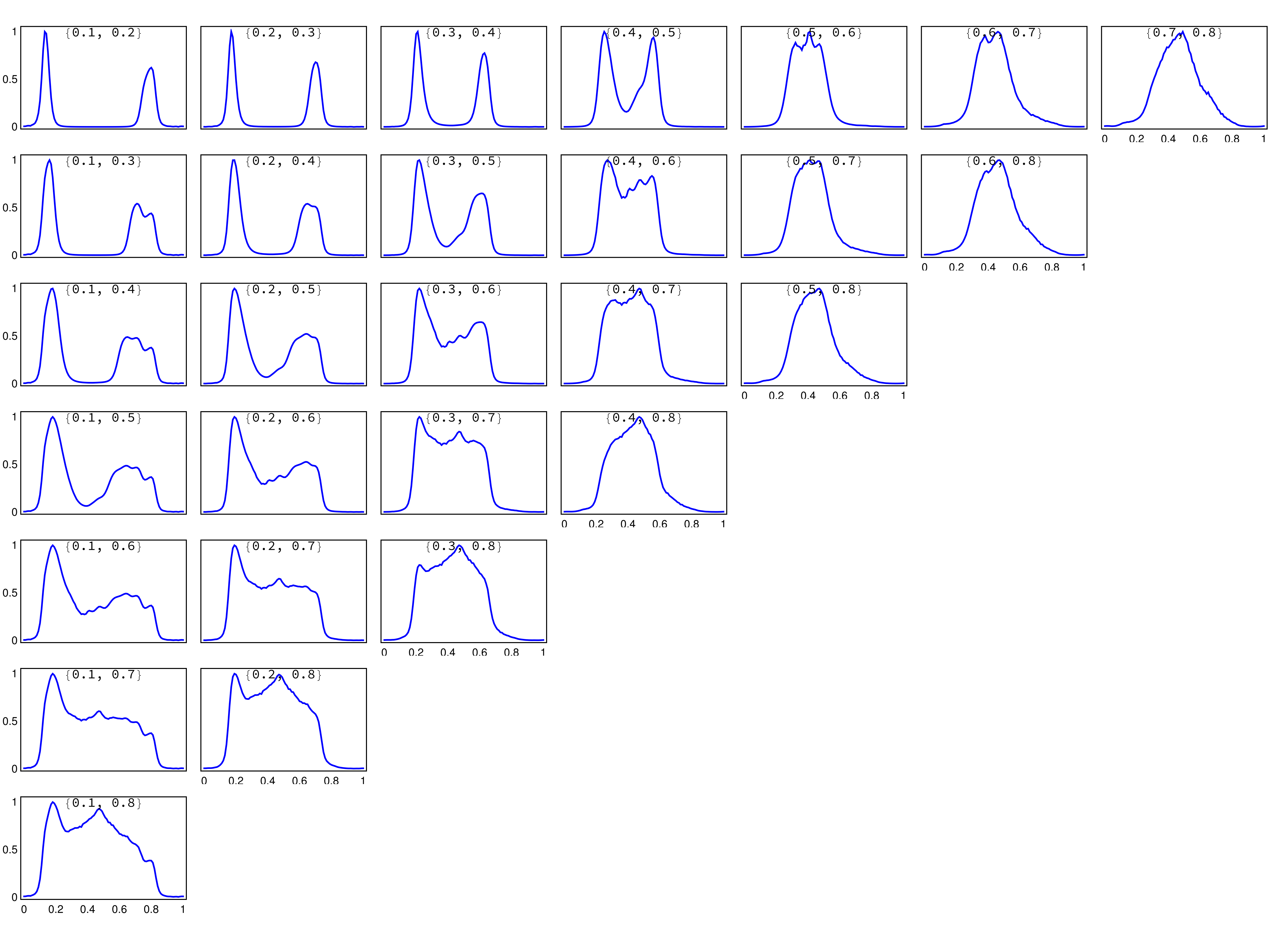}
	\caption{Evolution of the X-ray profile with altitude and extension of the emission region for $\rchi = 45\degr$ and $\zeta=44\degr$. The legend shows the lower and upper radius $\{r_{\rm in}/\rlight, r_{\rm out}/\rlight\}$, respectively denoted by $r_{\rm in}$ and $r_{\rm out}$, for the region of the separatrix emitting in X-ray. The upper rows show the individual light-curves that are progressively added together in the lower rows to the point where all are stacked to get only one possible light-curves as show by the lowest row.} 
	\label{fig:courbelumierexextension}
\end{figure*}

\begin{figure}[h]
	\centering
	\includegraphics[width=\linewidth]{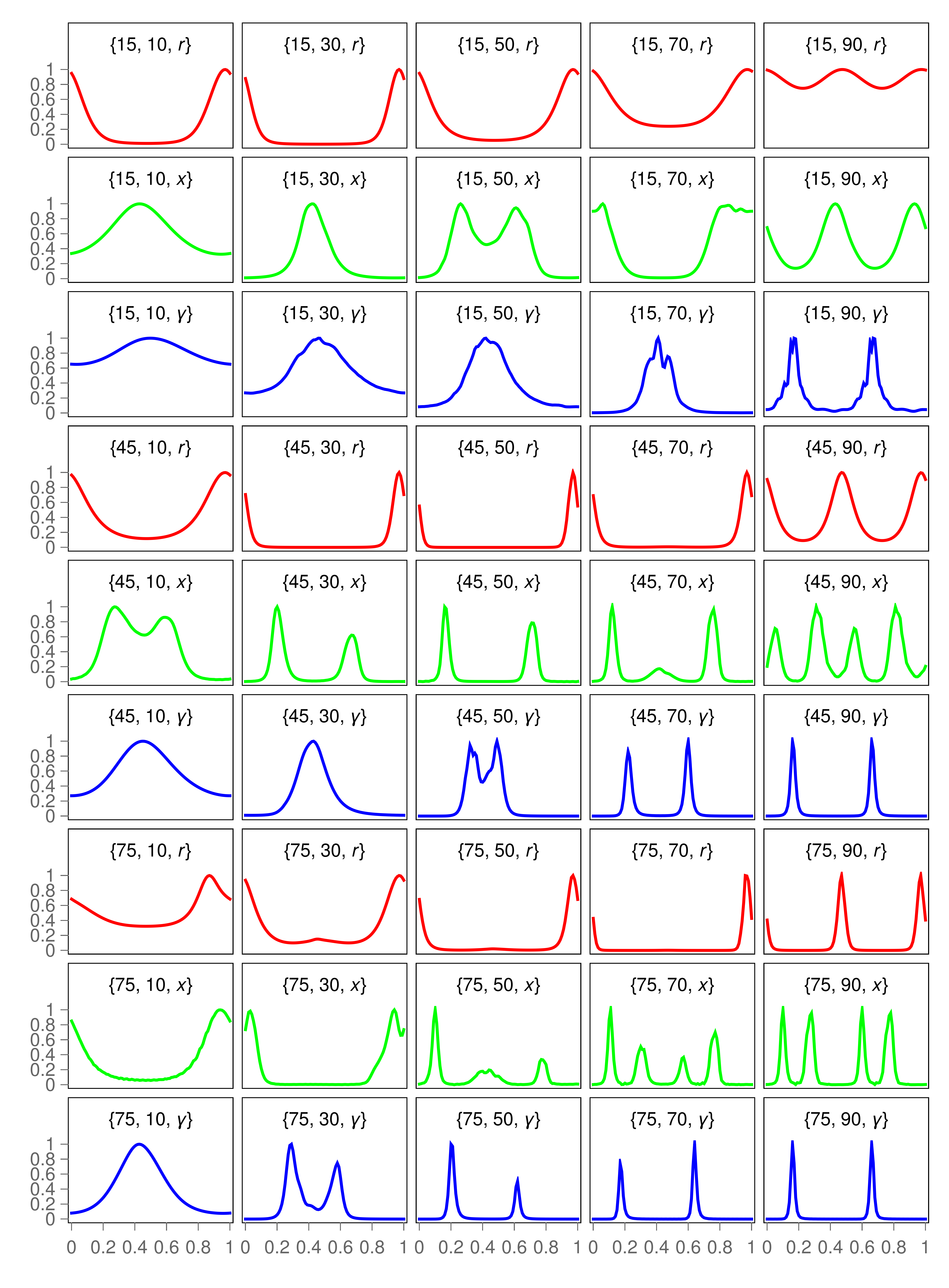}
	\caption{Multi-wavelength light-curves for $\rchi=15\degr$ on top panel, for $\rchi=45\degr$ on middle panel and for $\rchi=75\degr$ on bottom panel, with $\zeta=\{10\degr,30\degr,50\degr,70\degr,90\degr\}$. The legend is $\{\rchi, \zeta, em\}$ where $em$ stands for the three energy bands as $r$ for radio (red), $X$ for X-rays (green) or $\gamma$ for $\gamma$-rays (blue).}
\label{fig:multi_lamba}
\end{figure}

In this appendix we present several atlases of light-curves in radio, X-ray and $\gamma$-ray. Fig.~\ref{fig:atlas_radio_r0.2} shows a sample of typical radio pulse profiles when the emission height is set in the range $r/\rlight\in[0.1,0.11]$. Two radio pulses are observed for almost orthogonal rotators as expected. In these plots, even if the line of sight does not intersect the radio cone, we plot the pulse profiles, it corresponds to the bottom right and top left of the figure. The radio duty cycles appear therefore artificially wider because the radio visibility condition eq.~\eqref{eq:zeta_radio_visible_nord} or eq.~\eqref{eq:zeta_radio_visible_sud} is not met.
Fig.~\ref{fig:atlas_gamma_r1} shows the $\gamma$-ray light curves for different emission intervals, in blue for the range $r/\rlight\in[1,2]$, in green for the range $r/\rlight\in[2,3]$ and in red the total interval $r/\rlight\in[1,3]$. We observe a slight change in the profile shapes around the values satisfying $\rchi+\zeta \approx \pi/2$ corresponding to the line of sight grazing the current sheet. In all other geometries, the profile is insensitive to the distance.
Two examples of X-ray light curves are shown in fig.~\ref{fig:atlas_X_r0.2} for an emission height in the range $r/\rlight\in[0.1,0.2]$ and fig.~\ref{fig:atlas_X_r0.5} for an emission height in the range $r/\rlight\in[0.4,0.5]$.
As the extension of the emission region in X-ray is unconstrained, we plotted the effect of stacking together several adjacent emission sites to obtain a set of possible pulse profiles as shown in fig~\ref{fig:courbelumierexextension} for an obliquity of $\rchi = 45\degr$ and a line of sight inclination of $\zeta=44\degr$. Single and double peak structures are obtained with a symmetric or an asymmetric shape. Such profiles must be fitted to individual pulsars to extract useful information on the X-ray emission mechanisms.
Collecting the radio, X-ray and $\gamma$-ray pulse profiles, figure~\ref{fig:multi_lamba} shows the multi-wavelength light-curves for an obliquity $\rchi=15\degr$ and line of sight inclinations $\zeta=\{10\degr,30\degr,50\degr,70\degr,90\degr\}$ with the legend format $\{\rchi, \zeta, em\}$ where $em$ can be $r$ for radio, $X$ for X-rays or $\gamma$ for $\gamma$-rays, in red, green and blue solid lines respectively, in the first second and third row.
Other examples for $\rchi=45\degr$ and $\rchi=75\degr$ are also shown in the same figure~\ref{fig:multi_lamba}, middle panels and bottom panels respectively.

\begin{figure}[h]
	\centering
	\includegraphics[width=0.9\linewidth]{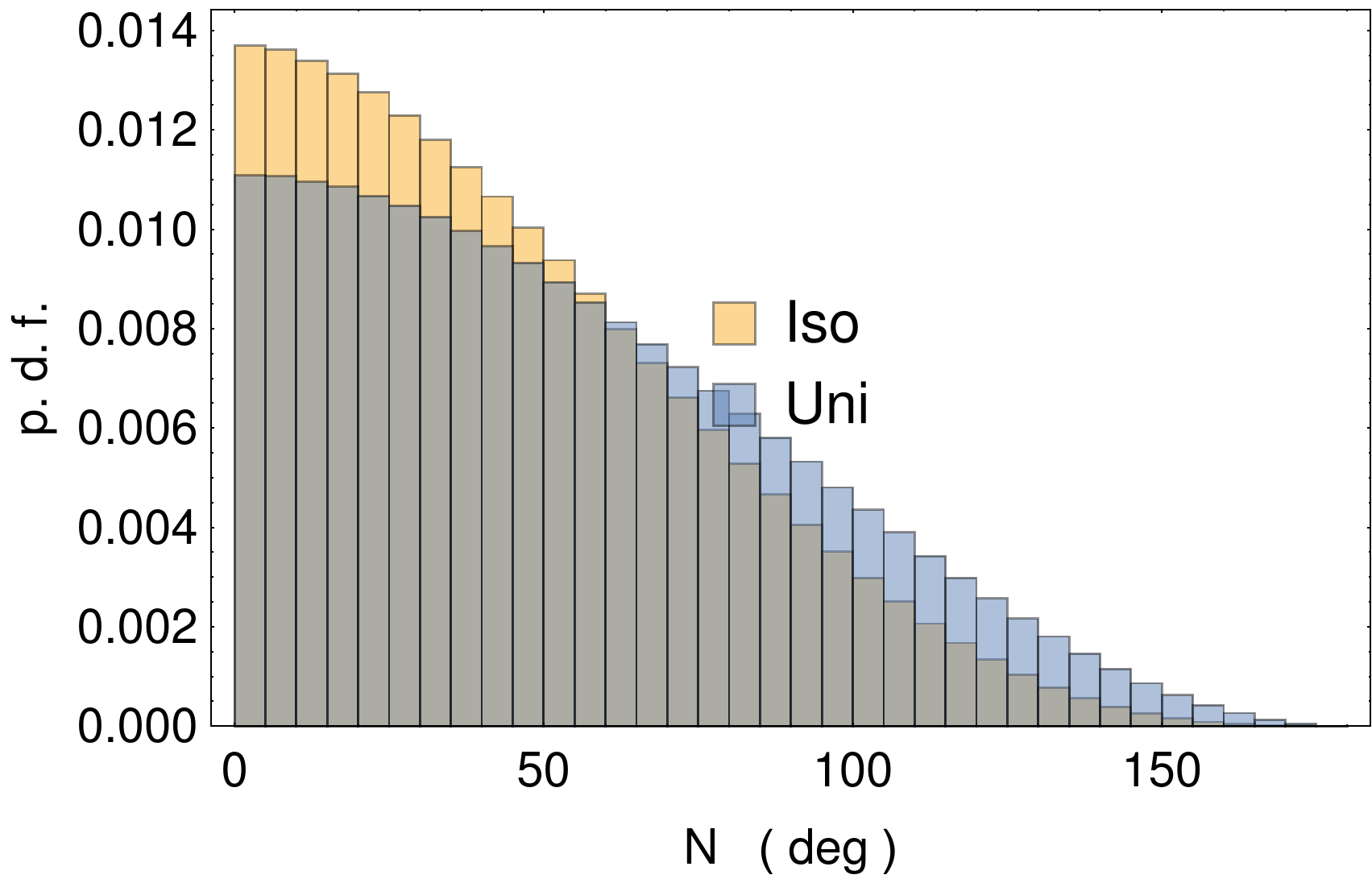}
	\caption{Probability density function of $N$ to decide for the pulsar radio visibility. The fraction of radio pulsars detected depends on the radio beam half-opening angle $\rho$.}
	\label{fig:visibilite_radio}
\end{figure}

\begin{figure}[h]
	\centering
	\includegraphics[width=0.9\linewidth]{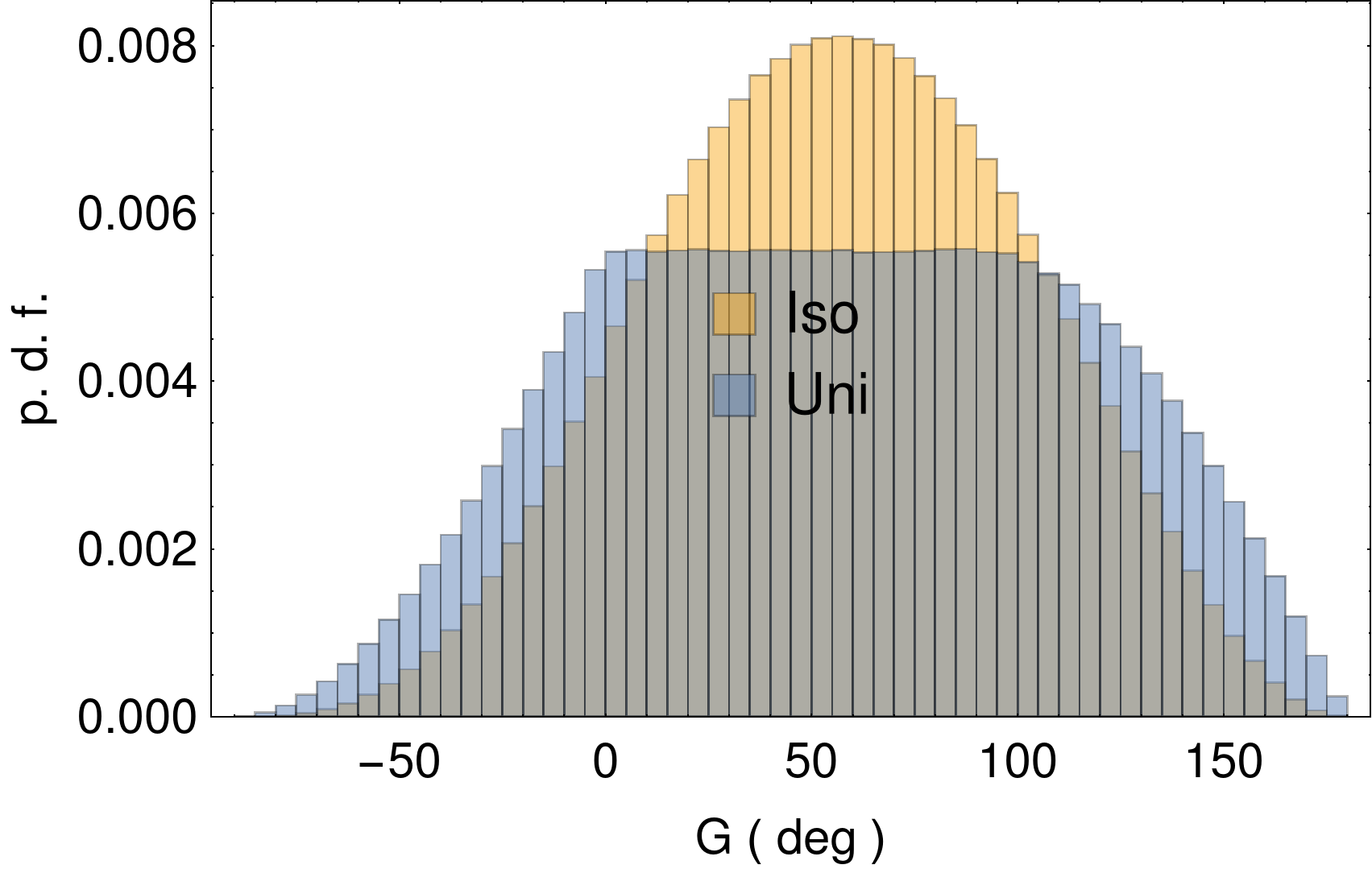}
	\caption{Probability density function of $G$ to decide whether the pulsar is visible in $\gamma$-ray or not. The region $G >0$ separates the pulsars visible in $\gamma$-ray from those invisible $G<0$. $G$ is given in degrees (deg).}
	\label{fig:visibilite_gamma}
\end{figure}

To decide whether a pulsar is visible in radio or $\gamma$-ray, we plotted the probability density functions (p.d.f.) for $N$ in fig.~\ref{fig:visibilite_radio} and for $G$ in fig.~\ref{fig:visibilite_gamma} for an isotropic and uniform distribution of magnetic obliquities, in orange and blue respectively.

\section{Dipolar coordinate system}
\label{app:A}

\begin{figure}[h]
	\centering
	\includegraphics[width=0.8\linewidth]{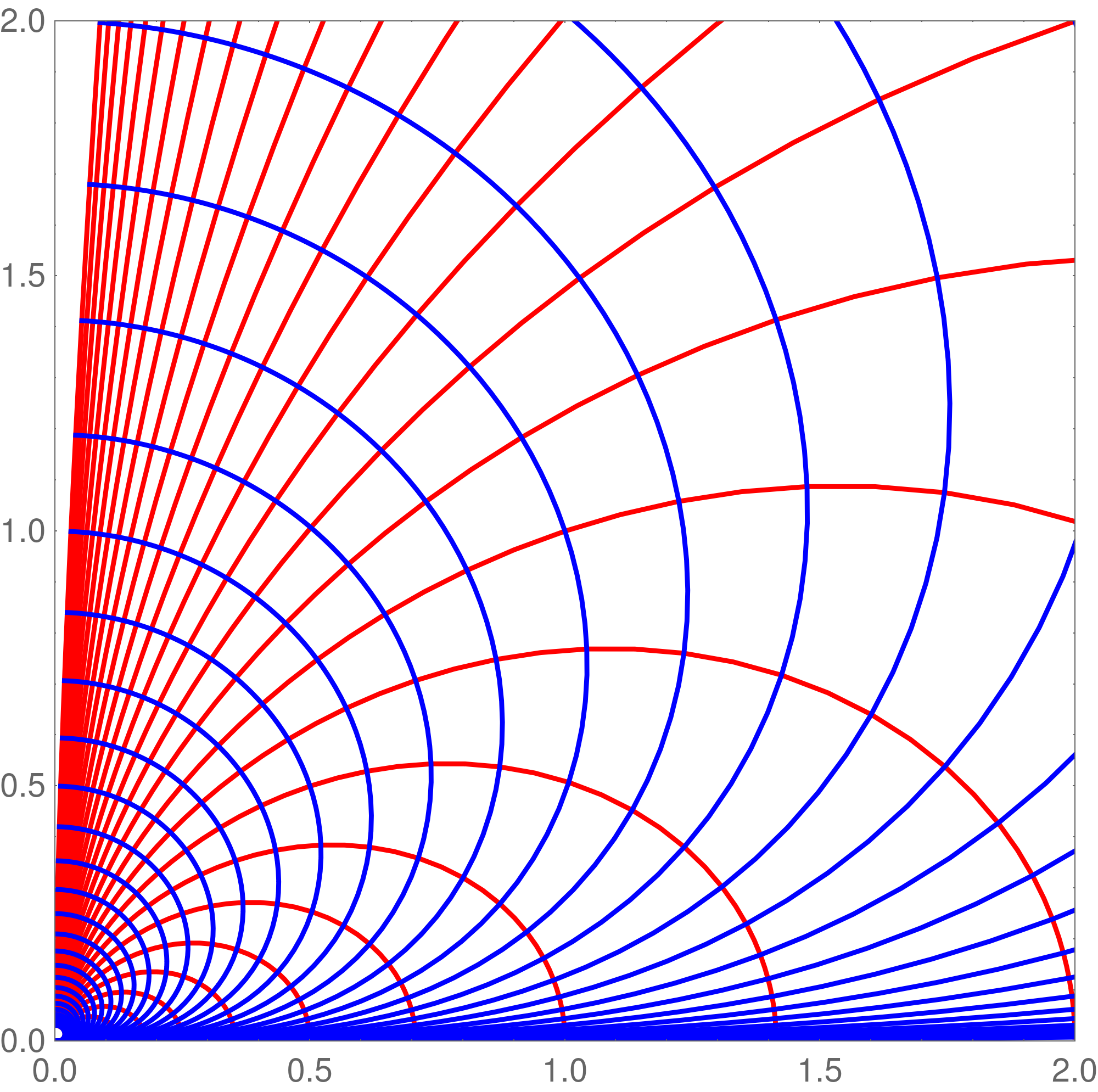}
	\caption{Dipolar curvilinear coordinate system showing lines of constant $\psi$, red lines, and lines of constant $\rho$, blue lines. These coordinates are orthogonal.}
	\label{fig:coordonneesdipolaires}
\end{figure}

Around a star whose field close to the surface is essentially dipolar, it is useful to choose a coordinate system following the geometry of the magnetic field lines. Starting from the spherical polar coordinates $(r,\theta,\phi)$ with natural basis $(\vec{e}_r, \vec{e}_\theta, \vec{e}_{\phi})$, we introduce the new coordinates $(\rho, \psi, \phi)$ and associated natural basis $(\vec{e}_\rho, \vec{e}_\psi, \vec{e}_{\phi})$ defined by
\begin{equation}
	\rho = \frac{\cos\theta}{r^2} \qquad ; \qquad \psi = \frac{r}{\sin^2\theta} \qquad ; \qquad \phi = \phi .
\end{equation}
The spherical radius~$r$ is found implicitly by solving the fourth order polynomial
\begin{equation}
	\rho^2 \, \psi \, r^4 + r - \psi = 0
\end{equation}
and the angle according to $\cos\theta = \rho \, r^2$. The coordinate system is shown in a meridional plane in fig.~\ref{fig:coordonneesdipolaires}.
The determinant of the Jacobian matrix
\begin{equation}
	J = \begin{pmatrix}
		\partial_r \rho & \partial_\theta \rho \\
		\partial_r \psi & \partial_\theta \psi 
	\end{pmatrix}
\end{equation}
is
\begin{equation}
	\textrm{det} J = \frac{\delta^2}{r^2 \, \sin^3\theta}
\end{equation}
with $\delta=\sqrt{3\,\cos^2\theta + 1}$.
The vectors of the natural basis are 
\begin{subequations}
	\begin{align}
		\vec{e}_\rho & = - \frac{2r^3\cos\theta}{\delta^2} \, \er - \frac{r^2\sin\theta}{\delta^2} \, \etheta \\
		\vec{e}_\psi & = \frac{\sin^4\theta}{\delta^2} \, \er - \frac{2\cos\theta\sin^3\theta}{r\,\delta^2}	\, \etheta
	\end{align}
\end{subequations}
and the inverse relations
\begin{subequations}
	\begin{align}
		\er & = - \frac{2\cos\theta}{r^3} \, \vec{e}_\rho + \frac{1}{\sin^2\theta} \, \vec{e}_\psi \\
		\etheta & = - \frac{\sin\theta}{r^2} \, \vec{e}_\rho - \frac{2r\cos\theta}{\sin^3\theta} \, \vec{e}_\psi .
	\end{align}
\end{subequations}
This basis is orthogonal and the metric~$g_{ik}$ diagonal with
\begin{subequations}
	\begin{align}
		g_{\rho\rho} & = h^2_{\rho\rho} = \frac{r^6}{\delta^2} \\
		g_{\psi\psi} & = h^2_{\psi\psi} = \frac{\sin^6 \theta}{\delta^2} \\
		g_{\phi\phi} & = h^2_{\phi\phi} = r^2 \, \sin^2\theta
	\end{align}
\end{subequations}
the surface element is
\begin{equation}
	dS_{ik} = \sqrt{g_{ii} \, g_{kk}} \, dx^i \, dx^k
\end{equation}
and the volume element
\begin{equation}
	dV = \frac{r^4 \, \sin^4\theta}{\delta^2} \, d\rho \, d\psi \, d\phi .
\end{equation}
Finally, defining the orthonormal basis relations between $(\vec{e}_{\hat{r}}, \vec{e}_{\hat{\theta}}, \vec{e}_{\hat{\phi}})$ and $(\vec{e}_{\hat{\rho}}, \vec{e}_{\hat{\psi}}, \vec{e}_{\hat{\phi}})$, we get
\begin{subequations}
	\begin{align}
		\vec{e}_{\hat{\rho}} & = - \frac{2\cos\theta}{\delta} \, \vec{e}_{\hat{r}} - \frac{\sin\theta}{\delta} \, \vec{e}_{\hat{\theta}} \\
		\vec{e}_{\hat{\psi}} & = \frac{\sin\theta}{\delta} \, \vec{e}_{\hat{r}} - \frac{2\cos\theta}{\delta} \, \vec{e}_{\hat{\theta}} \\
		\vec{e}_{\hat{\phi}} & = \vec{e}_{\hat{\phi}} .
	\end{align}
\end{subequations}
Particle are obliged to move along field lines defined by the separatrix surface. This corresponds to a motion with a fixed value for $\psi$ and $\phi$. This means that they move along the direction given by $\partial/\partial \rho$ thus along the unit vector $\vec{e}_{\hat{\rho}}$. Photons are emitted along the radially outward direction $\vec{t}$ such that $\vec{t} \cdot \vec{e}_{\hat{r}}>0$ implying 
\begin{equation}
	\vec{t} = 
	\begin{cases}
	- \vec{e}_{\hat{\rho}} & \textrm{ for } \theta \leq \pi/2 \\
	+ \vec{e}_{\hat{\rho}} & \textrm{ for } \theta > \pi/2 .
	\end{cases}
\end{equation}
For the aligned rotator, the separatrix is defined by $r_{\rm sep} = \rlight\,\sin^2\theta$ and emission starts at an altitude $h_{\rm e}$ thus $\sin \theta_{\rm e} = \sqrt{h_{\rm e}/\rlight}$.
A photon emitted at the position $(r_{\rm e}, \theta_{\rm e}, \phi_{\rm e})$ will propagate in a direction with colatitude $\theta_{\rm x}$ such that $\cos \theta_{\rm x} = t_{\rm z} = \vec{t} \cdot \ez$. Explicitly we find
\begin{equation}
	t_{\rm z} = \frac{3\cos^2\theta_{\rm e}-1}{\sqrt{3\cos^2\theta_{\rm e}+1}} = \frac{2-3\,h_{\rm e}/\rlight}{\sqrt{4-3\,h_{\rm e}/\rlight}} .
\end{equation}
Photons are produced between an emission height $h_1$ and $h_2>h_1$.
The observer will therefore detect X-rays if his line of sight $\zeta$ lies in the range spanned by the photon emission angle $\theta_{\rm x}$ i.e. $\zeta \in [\theta_{\rm x}^1,\theta_{\rm x}^2]$ with
\begin{equation}
	\cos \theta_{\rm x}^i = \sqrt{1-h_i/\rlight} \textrm{ with } i \in \{1,2\}.
\end{equation}
For an oblique dipole, we follow the same reasoning and rotate the Cartesian coordinate system to align the new $z$-axis with the magnetic moment axis, thus performing a rotation of angle~$\rchi$ along the $x$-axis with the rotation matrix
\begin{equation}
	R = 
	\begin{pmatrix}
		1 & 0 & 0 \\
		0 & \cos\rchi & \sin\rchi \\
		0 & -\sin\rchi & \cos\rchi
	\end{pmatrix} .
\end{equation}
The photon propagation direction in the upper hemisphere case becomes
\begin{equation}
	\delta \, \vec{t} = 3\cos\theta_{\rm e}\,\sin\theta_{\rm e} \, ( \cos \phi_{\rm e} \, \ex + \sin \phi_{\rm e} \, \ey) + (3\cos^2\theta_{\rm e}-1) \, \ez .
\end{equation}
The $z$ component is
\begin{equation}
	t_{\rm z} = \frac{(3\cos^2\theta_{\rm e}-1)\,\cos\rchi - 3\cos\theta_{\rm e}\,\sin\theta_{\rm e} \, \sin \phi_{\rm e} \, \sin\rchi}{\sqrt{3\cos^2\theta_{\rm e}+1}}.
\end{equation}
The lower and upper bound for this component are
\begin{subequations}
	\begin{align}
	t_{\rm z}^\pm(\theta_{\rm e}, \rchi) 
	& = \frac{3 \cos\theta_{\rm e} \, \cos ( \theta_{\rm e} \pm \rchi) - \cos \rchi}{ \sqrt{3\cos^2\theta_{\rm e}+1} } \\
	& = \cos( \rchi \pm \arccos t_{\rm z}^\pm(\theta_{\rm e}, 0) ) = \cos ( \rchi \pm \theta_{\rm x} ).
	\end{align}
\end{subequations}
Consequently, the photon emission direction when starting at the position $\theta_{\rm e}$ for an oblique rotator with obliquity~$\rchi$ stays between the two extremal values
\begin{equation}
	\theta_{\rm x}^ \pm(\rchi) = \rchi \pm \theta_{\rm x}(0) .
\end{equation}
This formula was expected since the X-ray half-opening angle is $\theta_{\rm x}$ and the cone is centred around the magnetic axis.
Note that this definition is exactly the same as for the radio emission cone opening angle namely
\begin{equation}
	\theta_{\rm x}(0) = \theta_{\rm e} + \arctan \left( \frac{\tan \theta_{\rm e}}{2} \right) .
\end{equation}
The X-ray visibility will therefore mimic the radio visibility condition, the only difference being the size of the emitting cone because of the varying emission altitude postulated in X-rays. We emphasize that this view is correct for a static dipole and thus restricted to altitudes of at most a fraction of the light-cylinder. High altitude emission will significantly deviate from the above guess. 

\end{document}